\documentclass[conference,10pt]{IEEEtran}
\usepackage{cite}
\usepackage{amsmath,amssymb,amsfonts}
\usepackage{algorithmic}
\usepackage{graphicx}
\usepackage{textcomp}
\usepackage{xcolor}
\usepackage[hyphens]{url}
\usepackage{balance}

\def\BibTeX{{\rm B\kern-.05em{\sc i\kern-.025em b}\kern-.08em
    T\kern-.1667em\lower.7ex\hbox{E}\kern-.125emX}}

\usepackage[linesnumbered]{algorithm2e}
\usepackage{soul}
\usepackage{setspace}
\usepackage{subfigure}
\SetAlCapSkip{0pt}
\usepackage{etoolbox}
\newbool{inccomment}
\setbool{inccomment}{true}
\newcommand{\sms}[1]{\ifbool{inccomment}{{\color{blue}SMS: #1}}{}}
\newcommand{\akj}[1]{\ifbool{inccomment}{{\color{magenta}AKJ: #1}}{}}
\newcommand{\sjl}[1]{\ifbool{inccomment}{{\color{red}SJL: #1}}{}}
\newcommand{\sbo}[1]{\ifbool{inccomment}{{\color{cyan}SBO: #1}}{}}

\newcommand{\superscript}[1]{\ensuremath{^{\textrm{#1}}}}

\def\wg{\superscript{\dag}}
\def\wu{\superscript{\S}}
\title{\huge Virtual Coset Coding for Encrypted Non-Volatile Memories with Multi-Level Cells} 
\author{\IEEEauthorblockN{Stephen Longofono\wg, Seyed Mohammad Seyedzadeh\wu, Alex K. Jones\wg\\}
\IEEEauthorblockA{\wg Electrical and Computer Engineering Department, University of Pittsburgh, \wu AMD Research, USA\\
\normalsize{Email: stl77@pitt.edu, Mohammad.Seyedzadeh@amd.com, akjones@pitt.edu}\vspace{-0.5em}
}
}

\begin{document}
\maketitle
\thispagestyle{plain}
\pagestyle{plain}


\begin{abstract}
    Recently, Phase-Change Memory (PCM) has become a popular commercialized non-volatile memory (NVM), which has been deployed as a backing memory for DRAM main memory, secondary storage, or even as a DRAM main memory replacement.   Like other NVMs, PCM has asymmetric access energy; writes dominate reads.  When considering multi-level cells (MLC), this asymmetry can vary by an order of magnitude.  Many schemes have been developed to take advantage of the asymmetric patterns of `0's and `1's in the data to reduce write energy.  Because the memory is non-volatile, data can be recovered via physical attack or across system reboot cycles. To protect information stored in PCM against these attacks requires encryption.  Unfortunately, most encryption algorithms scramble `0's and `1's in the data, effectively removing any patterns and negatively impacting schemes that leverage data bias and similarity to reduce write energy.  In this paper, we introduce Virtual Coset Coding (VCC) as a workload-independent approach that reduces costly symbol transitions for storing encrypted data. VCC is based on two ideas.  First, using coset encoding with random coset candidates, it is possible to effectively reduce the frequency of costly bit/symbol transitions when writing encrypted data.  Second, a small set of random substrings can be used to achieve the same encoding efficiency as a large number of random coset candidates, but at a much lower encoding/decoding cost. Additionally, we demonstrate how VCC can be leveraged for energy reduction in combination with fault-mitigation and fault-tolerance to dramatically  increase the lifetimes of endurance-limited NVMs, such as PCM. We evaluate the design of VCC and demonstrate that it can be implemented on-chip with only a nominal area overhead. VCC reduces dynamic energy by 22-28\% while maintaining the same performance.  Using our multi-objective optimization approach achieves at least a 36\% improvement in lifetime over the state-of-the-art and at least a 50\% improvement in lifetime vs. an unencoded memory, while maintaining its energy savings and system performance.
\end{abstract}

\section{Introduction}
\label{sec:introduction}
Next generation server architectures leverage tiered main memory combining conventional DRAM and dense non-volatile memory (NVM). Such tiered memory systems both support increasingly large application footprints and, in particular, reduce energy consumption, especially static energy, of main memory
~\cite{NVM_Facebook}.  
Products like the Intel Optane DC are demonstrating commercial approaches of using phase-change memory (PCM) in tiered memory Cascade Lake servers~\cite{OPTANE_DC}.

Although tiered-memory systems are a promising solution, the limited endurance of NVM presents challenges.  A typical PCM cell can tolerate on the order of $10^8$ writes~\cite{xue2011emerging} before failing to reliably switch.  Once the lifetime of a PCM cell is reached, it will become ``stuck'' in its present state.  Thereafter, the cell is immutable, but its value can still be read. In order to realize the cost savings of incorporating PCM as main memory for a server, memory lifetime must be addressed.  Existing techniques such as differential write~\cite{PCM_Stats} and flip-n-write~\cite{cho2009flip} leverage data similarity to reduce the amount of cell switches or by encoding data before writing it back from the last-level cache to better match the existing cell state.  

Unfortunately, as shown in Section~\ref{sec:background-cosets}, for encrypted data these techniques are insufficient to reduce writes.  Encrypted data will be indistinguishable from random numbers; any patterns present in the unencrypted data are not preserved.  If a cell is equally likely to store a `0' or a `1', then we expect half of the cells in any given write will require changing the cell state, on average which increases energy and degrades reliability~\cite{5544298}.  
In this scenario, coset coding techniques~\cite{forney1988coset} can transform data to better match  existing cell values.  By selecting a large enough set of codes, the likelihood that at least one of the transformations will match the existing data is improved.  However, this presents a tradeoff in both performance and area; the storage, energy, and latency penalties increase substantially with the number of codes tested.

To maintain a large number of codes while addressing the energy and latency penalty of encoding, we propose Virtual Coset Coding (VCC).  
Standard cosets store or generate codes for the entire data block (typically a cache line or memory row) and test the transformation from each candidate against the existing data.  Instead, VCC \textit{builds and evaluates} many coset transformations in parallel using a few much smaller random \textit{kernel codes}.  Each kernel and its inversion are tested on partitions of the data block in parallel and the minimum of the two are concatenated together to form the best candidate based on that kernel.  VCC, using a small number of kernels in this fashion both allows rapid evaluation of the encoded data while achieving a sufficiently random-like set of ``virtual'' coset candidates that can achieve similar benefits to the static random coset candidates. 
We study VCC as applied to MLC PCM which has been encrypted by counter-mode AES~\cite{NIST-counter-mode}. Specifically, this paper makes the following contributions:
\begin{itemize}
    \vspace{-0.1in}
    \item We present Virtual Coset Coding, a technique for generating and evaluating pseudo-random codewords for data transformation with reduced computational complexity.
    \item We present techniques for reducing write energy in SLC and MLC phase-change memory using encrypted data while opportunistically mitigating failed cells.
    \item We present an implementation to support the use of VCC at the memory controller, along with a method to generate random codewords from encrypted data.
    \item We present a detailed analysis of the impacts of VCC and the supporting architecture in terms of PCM lifetime and dynamic energy for memory-intensive benchmarks.
\end{itemize}

\section{Background}
\label{sec:background}
In this section we provide a background on PCM and its endurance-related failure modes, alongside related error correction and encoding techniques which have been used to mitigate transient and permanent faults.

\subsection{Phase-change Memory Characteristics}
\label{sec:background-pcm}
PCM technology is a resistive memory which interprets the resistance measured across a cell as logical `1' (SET) in the low-resistance state or logical `0' (RESET) in the high-resistance state.  PCM cells are formed from chalcogenide materials such as $Ge_{2}Sb_{2}Te_{5}$, which realize a change in conductance when the material is in an amorphous state vs. its crystalline lattice state~\cite{PCM_Stats}.  In order to modify the state of a cell, a voltage greater than the material's switching threshold is applied, inducing current to flow through the cell and raising its temperature via joule heating.  When the phase-change material in the cell reaches its melting temperature (~$600^{\circ}C$) it changes to its amorphous state, where it can be held by abruptly removing the potential across the cell.  If instead the cell is continually heated above its crystallization temperature, the cell material tends to relax into its low-resistance state.  In practice, this is achieved by applying a short pulse with high current to melt the material, and a series of longer, lower-current pulses to allow the material to crystallize~\cite{PCM_Background}.

The range of resistances of the cell material is logically divided into regions which are used to determine the cell state when a cell is accessed.  In a single-level cell (SLC), the resistance range is divided into two regions, which represent the two states that each data symbol represents.  Since the physical state of the PCM material is not discrete, it can be divided into smaller regions to represent a symbol with more states.  
Such multi-level cells (MLC) have a smaller noise margin for both sensing and programming but have the potential to multiply the capacity.  In the context of this work, we consider the case of MLC where there are four distinct levels per cell, sufficient to encode two logical bits.

Seminal studies of PCM reliability identified two main failure modes: a failure mode which results in a permanent low-resistance state (``stuck-at 1'') and another which results in a permanent high-resistance state (``stuck-at 0'')~\cite{PCM_WEAR}.  The former is characterized by a cell failing to reach a high-resistance state under increased programming time and current, suggesting that the switching threshold was no longer permitting adequate joule heating to melt the material.  Using energy-dispersive spectroscopy techniques, the authors observed a depletion of $Ge$ in the phase-change material, suggesting that material ``phase-segregation'' was responsible for the loss of cell programmability.  The latter failure mode is characterized by a cell exhibiting a substantially higher resistance than either of its nominal states.  This failure mode was attributed to a physical separation (void) at the interface of the phase-change material and its access circuit, which tended to increase with repeated writes to the cell~\cite{PCM_WEAR}.  Using fabrication studies of DRAM memory~\cite{seyedzadeh2017mitigating,seyedzadeh2017counter,seyedzadeh2018mitigating} as an indication of the PCM manufacturing process, we expect that some cells will wear faster than others due to process variation~\cite{Process_Variation}, and that there are spatial correlations among those cells such that they are more likely to cluster together in the same memory row~\cite{yuan2011yield}.

The general consensus among PCM reliability studies is that the extremes of temperature are the primary cause of cell wear~\cite{PCM_WEAR,RELIABILITY_STUDY,SET_TEMPERATURE_ONLY,SET_FAIL_STUDY}. \textit{Thus, reducing write energy simultaneously improves energy efficiency and prolongs cell lifetime.}  In turn, techniques designed to increase PCM lifetime can target reducing writes overall, or alternatively reducing those writes which require more energy transfer to the cell media.  In the case of SLC, this is one and the same with the RESET state. For MLC, the mapping of symbol values to cell states dictates which states should be reduced. However, the smaller noise margins and greater uncertainty involved with programming MLCs require additional programming steps that can accelerate cell wear.

Programming a PCM cell is probabilistic, in the sense that after a pulse is applied to program a cell, it will change its state at a non-deterministic rate.  Individual cells change state at different rates relative to one another, but the same access circuit is used to program them all.  Variations in the initial state, pulse time, or voltage applied will also influence the distribution of the change of resistance.  In prototype MLC devices, it was observed that applying a long SET pulse followed by a RESET pulse was necessary prior to programming to the intermediate states; this procedure provides a more reliable starting point, reducing the possibility of under- or over-programming the cell~\cite{MLC_DEVICE}.  Unfortunately, that means that it is not feasible to move to the intermediate states directly; a full SET and RESET are required, followed by a program-and-check sequence to achieve the desired state.  These intermediate states require up to an order of magnitude more energy to program, and we expect those states to accelerate cell wear.

Reducing wear is an important preventative measure if PCM and other endurance-limited memories are to be used in main memory.  However, cells will inevitably fail, and the resultant faults must be addressed.  General error correction techniques are widely employed in memories to identify and mitigate transient faults.  In the context of PCM, they can also be employed to identify and mitigate permanent faults associated with memory cell wear.  In the next section, we discuss the relevant error correction techniques employed for main memory.

\subsection{Error Correction Techniques}
\label{sec:background-error-correction}
A commonly used error correction technique for main memory is Hamming's error correcting code (ECC)~\cite{hamming1950error}.  Paired to the word size of a modern computer, each 64 bits of data is protected by 8 parity bits, for a total of 72 bits per code word.  This ECC scheme permits single-error correction, and double-error detection (SECDED ECC).  For fault rates less than $10^{-6}$, such fault correction is often sufficient for main memory~\cite{nair2013archshield}.  However, if process variation results in a cluster of several faults in the same row, or an aging memory exceeds this raw fault rate, SECDED ECC will fail.

An alternative technique which allows for greater fault protection is error-correcting pointers (ECP)~\cite{ECP}.  A memory with ECP stores a pointer for each row (length $log_{2}(|row|)$) alongside a replacement bit.  When a faulty cell is identified, its position in the row is encoded in the pointer, and its correct value is written to the replacement bit.  Using $N$ such pointer and replacement bit pairs, up to $N$ faults can be mitigated per row (ECPN).  Such a scheme improves on SECDED, but can have high overhead ({\raise.17ex\hbox{$\scriptstyle\mathtt{\sim}$}}10 bits per fault protected) and is inefficient if faults occur within the ECP pointers.

\subsection{Coset Encoding Techniques}
\label{sec:background-cosets}
Cosets are a class of codes which are used to transform data, in contrast with ECC and ECP which mitigate faults directly.  Cosets transform data being written to memory such that the new value matches what is already there, with the goal of reducing bit changes.  In the context of a memory with permanent faults, the same approach can be used to match the value of faulty cells, effectively masking any errors associated with the faulty cells.  At the same time, such techniques can be used to optimize for other constraints such as minimizing bit changes or write energy.  If there are multiple cosets which mask a fault, then a secondary optimization can be used to select the best coset.  These schemes are subject to a ``read-modify-write'' overhead that each write requires a read to retrieve the original data over the memory bus for comparison, for which some proposed solutions are emerging~\cite{8792091}.

Data block inversion (DBI)~\cite{bae200880,hollis2009data}, originally for data transmission over bus lines,  writes the data itself when it would result in less than or equal to half of the bits changing.  Otherwise it writes the inverted data, encoded by a single bit per data block.  DBI is effective when data exhibits regular patterns, such as integers; integers used in computation typically do not change the most significant bits (MSBs) as often as the least significant bits, so it is possible to reduce bit changes 
by inverting negative numbers stored in twos-complement form.  Simply dividing the word size into two data blocks can capture this pattern and incur fewer writes on the MSBs of the memory.  As the size of the block is reduced, more fine-grained patterns can be captured and with more flexible application, at the expense of additional auxiliary bits to encode which blocks are inverted.

Flip-N-Write (FNW)~\cite{cho2009flip} is a scheme to reduce the number of bits written in memory by selectively inverting sub-blocks of data. In general, FNW divides the data into blocks of $k$ bits, and writes each block either directly or inverted, whichever minimizes the number of written bits. It uses one overhead bit per block to track the encoding and allow retrieval of the original data. More formally, for encoding an n-bits block, $D$, FNW can be viewed as using two coset candidates, $V_{0}=`0,...,0$' and $V_{1}=`1,...,1$', to encode $D$ as either $C = V_0 \oplus D$ or $C =V_1 \oplus D$, whichever leads to a lower cost\footnote{The symbol $\oplus$ denotes the \texttt{XOR} (exclusive OR) operation.}. Of course, the index, $i$ of the coset candidate used for encoding, which is either 0 or 1, should be sent to the decoder to allow the retrieval of the data as $D = V_i \oplus C$.

Flipcy~\cite{Flipcy}, similar to DBI, writes either the data, the 1's complement of the data, or the 2's complement of the data to reduce error-prone or computationally expensive symbols in MLC PCM.  This technique originally targeted triple-level cells with otherwise unacceptable programming error rates but can also be applied to more reliable single- and double-level cells.  In the same way as with DBI and FNW, Flipcy can optimize for reliability or energy.
%
%
Restricted coset coding \cite{seyedzadeh2018enabling} opens a small amount of space within each cache block via lightweight compression to store the auxiliary bits of coset encoding, thereby lowering encoding data block granularity and optimizing the cost function. 

A two-dimensional FNW in \cite{maddah2015cafo} alternates row and column FNW in an iterative fashion until no more Hamming weight reduction can be achieved. In contrast, XORTransfer~\cite{leereducing} exploits inter-sub-block data similarity in data blocks via zero data remapping and the universal base technique.

Unfortunately, the effectiveness of these coset techniques is degraded when applied to encrypted workloads.  In the next section, we discuss why the randomizing effect of encryption is problematic, existing techniques which account for unbiased data, and motivate the need for our VCC technique. 

\section{Motivation}
\label{sec:motivation}
All of the approaches discussed in Section~\ref{sec:background-cosets} are well designed for unencrypted workloads and take advantage of the 1/0 bias in the data. When the workloads are encrypted or have random patterns, encodings that rely on biased coset candidates lose their effectiveness since there is no longer biased data. As we will demonstrate in this section, random coset candidates are more effective for encoding random data.

Typically, Coset Coding~(CC)~\cite{forney1988coset} acts in two steps. First, it \texttt{XOR}s the data block, $D$, with $N$ coset candidates, $V_{0}, ... ,V_{N-1}$ that results in $N$ encoded candidates, $X_{0}, ..., X_{N-1}$. Second, it selects the optimum candidate $X_{opt}$ that optimizes the stated objective. Note that if the data block, $D$, is $n$-bits long, then each coset candidate is also $n$-bits long. However, the bits of the coset candidates can be selected to exhibit either a biased pattern~\cite{cho2009flip} or a random~\cite{jacobvitz2013coset} pattern. 

First, we consider the case of random coset encoding (RCC), where the coset candidates $V_{0}...V_{N-1}$ are random and independent.  The encoded candidates, $X_{0}... X_{N-1}$, in this case are also random and independent. Since each data bit is equally likely to be `0' or `1', and each coset bit has the same probability, then the probability of any individual bit changing is 0.5.  Using the binomial distribution, we can derive the probability of at least $m$ bits changing by summing over $Binomial(n,i)$, $i\in [0,m]$, and subtracting the resultant value from one.  Finally, to express the expected number of changed bits under $N$ cosets, we take use the discrete version of the identity $E[X] = \sum_{n=0}^{\infty} P(X>n)$ to derive:

%
\label{eq:1}
\begin{equation}
E_{RCC}=\sum^{n-1}_{m=0}\bigg(1-\sum^{m}_{i=0}\binom{n}{i}p^{i}(1-p)^{n-i}\bigg)^N
\end{equation}

\noindent where $m$ indexes the number of possible bit changes in a block of length $n$ and the argument of the outer summation represents the probability that $N$ applied cosets resulted in at least $m$ bits changing. Since  $log_{2}(N)$ bits are needed to identify the optimum coset candidate among the $N$ coset candidates, the average number of `1's in the encoded data block is $E_{RCC}$ + $\frac{\log_{2}(N)}{2}$.

Next, we consider the case of biased coset coding (BCC), where each word of length $n$ is divided into $k=log_{2}(N)$ sections of $\frac{n}{k}$ bits each.  If each subsection is written back unchanged or inverted, there are two outcomes per section, for a total of $2^{log_{2}(N)}=N$ possible coset candidates.  Using the expression for the expected number of bit changes per block~\cite{cho2009flip}, multiplied by $k$ blocks per word, we derive the expected number of bit changes for BCC as described above:

\begin{equation}
    \label{eq:2}
    \vspace{-0.1in}
    \begin{split}
    E_{BCC} &= k \bigg( 
    \sum_{i=0}^{\frac{n}{2k}} \frac{i}{2^{\frac{n}{k}+1}} \binom{\frac{n}{k}+1}{i}
    +\\
    &\sum_{i=\frac{n}{2k}+1}^{\frac{n}{k}+1}  \frac{(\frac{n}{k}+1-i)}{2^{\frac{n}{k}+1}}\binom{\frac{n}{k}+1}{i}
    \bigg)
    \end{split}
    \vspace{-0.1in}
\end{equation}

%
\noindent{}

%
Using these expressions, we can compare the effectiveness of RCC and BCC vs. writing the unencoded data block for random data.
\begin{figure}[!t]
    \centering
    \includegraphics[width=0.475\textwidth]{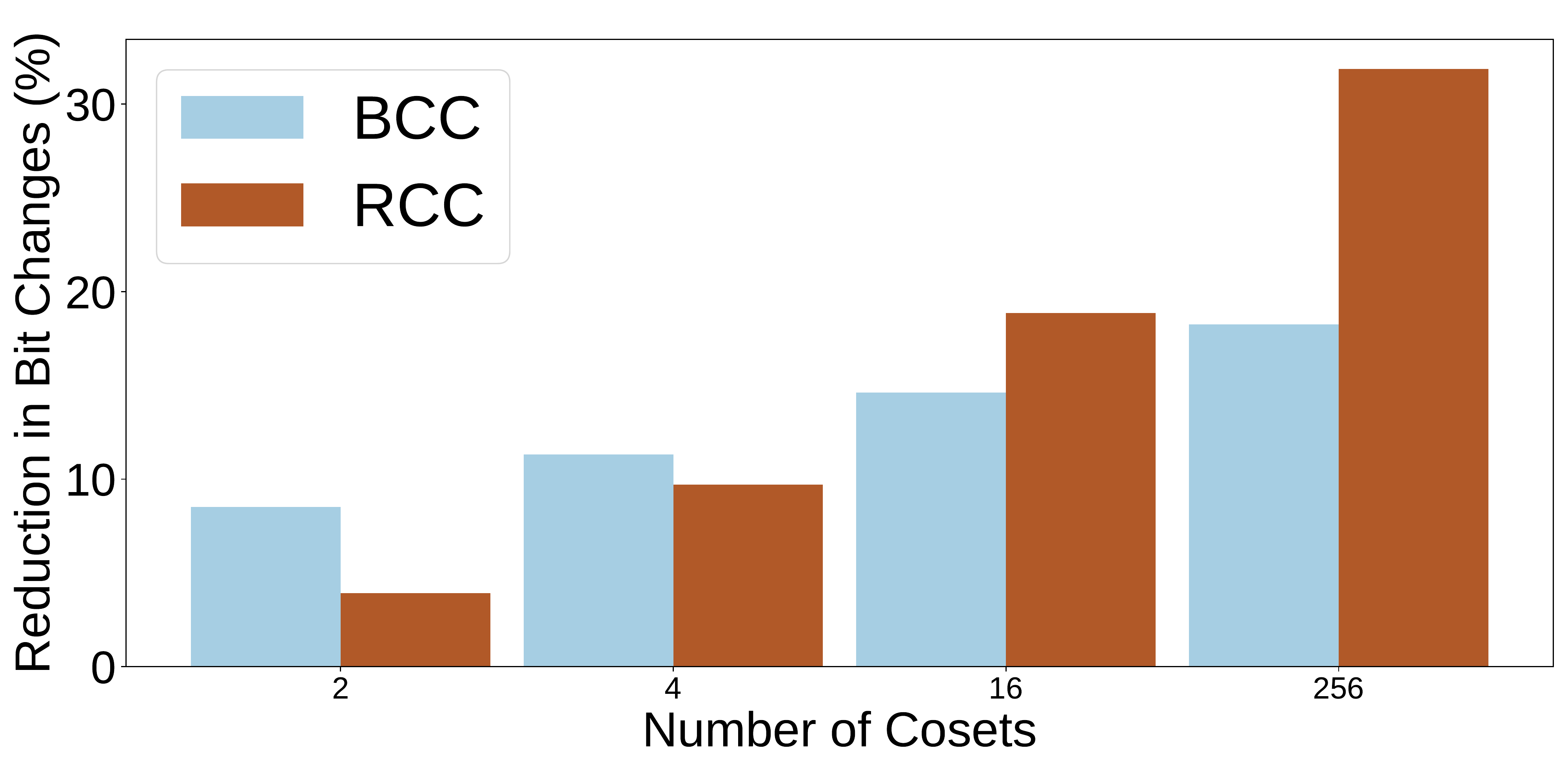}
    \vspace{-0.1in}
    \caption{The reduction in changed bits achieved by RCC and BCC for random data blocks, relative to writing the unencoded block.}
    \label{fig:bit_reduction}
    \vspace{-0.2in}

\end{figure}
Fig.~\ref{fig:bit_reduction} shows the reduction of changed bits achieved by  
RCC and BCC, for random data blocks relative to the unencoded data block\footnote{The expected number of changes in a random block of $n$ bits is $\frac{n}{2}$.}.  The figure shows the data block granularity fixed at $n=64$ when $N$ = 2, 4, 16, and 256, respectively.  We can see that with a small number of codes, BCC is more effective at reducing bit changes.  As the number of codes increases to 16 and 256, RCC becomes more effective, outperforming BCC by a considerable margin for the $N=256$ case. Thus random cosets perform better than biased cosets on encrypted (\textit{i.e.,} unbiased) data~\cite{Hybrid_Coset}.  

\begin{figure}[!t]
    \centering
    \includegraphics[width=0.475\textwidth]{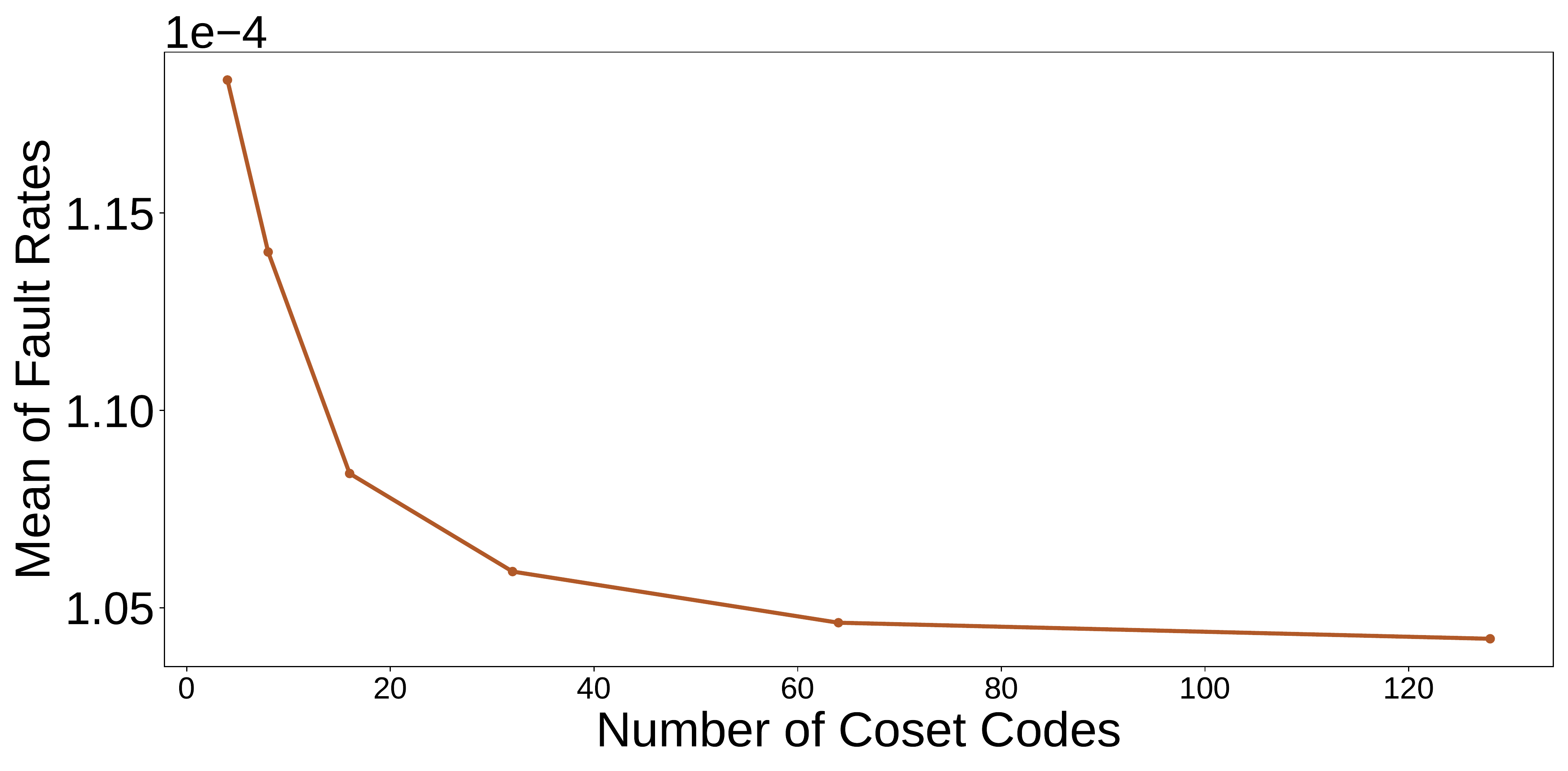}
    \vspace{-0.1in}
    \caption{Improvement in observed fault rates when coset codes are applied to mask faulty bits, using a nominal fault rate of $10^{-2}$ and benchmarks from the SPEC CPU\textsuperscript{\textregistered} 2017 benchmark suite~\cite{SPEC2017}.}
    \label{fig:coset_sweep}
\vspace{-0.2in}
\end{figure}

Since it is desirable to both reduce bit changes and mitigate the effects of failed cells, we evaluate how well $N$ cosets can mitigate errors associated with wear-induced faults.  Fig.~\ref{fig:coset_sweep} shows the results of a simulation 
using a pre-generated fault map with a fault rate of $10^{-2}$ to model a memory with an extreme fault rate.  For each write to an address with one or more faults, one of the $N$ random cosets is applied via \texttt{XOR} to best match the data being written to the existing data at the same address and mitigate errors associated with the faults.  We can see that as the number of cosets $N$ increases, the mean observed bit-error rate is reduced. Note that in order to decide that a fault is masked, it is necessary to identify and track faults at runtime.  Several fault repositories have been proposed for efficiently tracking faults up to fault rates approaching $10^{-2}$~\cite{nair2013archshield,FLOWER}.  For the remainder of the paper, we assume some such mechanism is in place, and its impact is identical on all techniques under test.

We have demonstrated that both BCC and RCC can be effective at reducing cell wear but in both cases, there is a trade space between the granularity at which we apply cosets, the number of codes used, the latency associated with evaluating the cosets, and the amount of auxiliary information that must be stored.  
The next section presents the VCC concept and discuss how it improves upon RCC and BCC in terms of auxiliary overhead, runtime latency, and optimization goals.

\begin{figure*}[!ht]
\begin{center}
\includegraphics[width=0.8\textwidth]{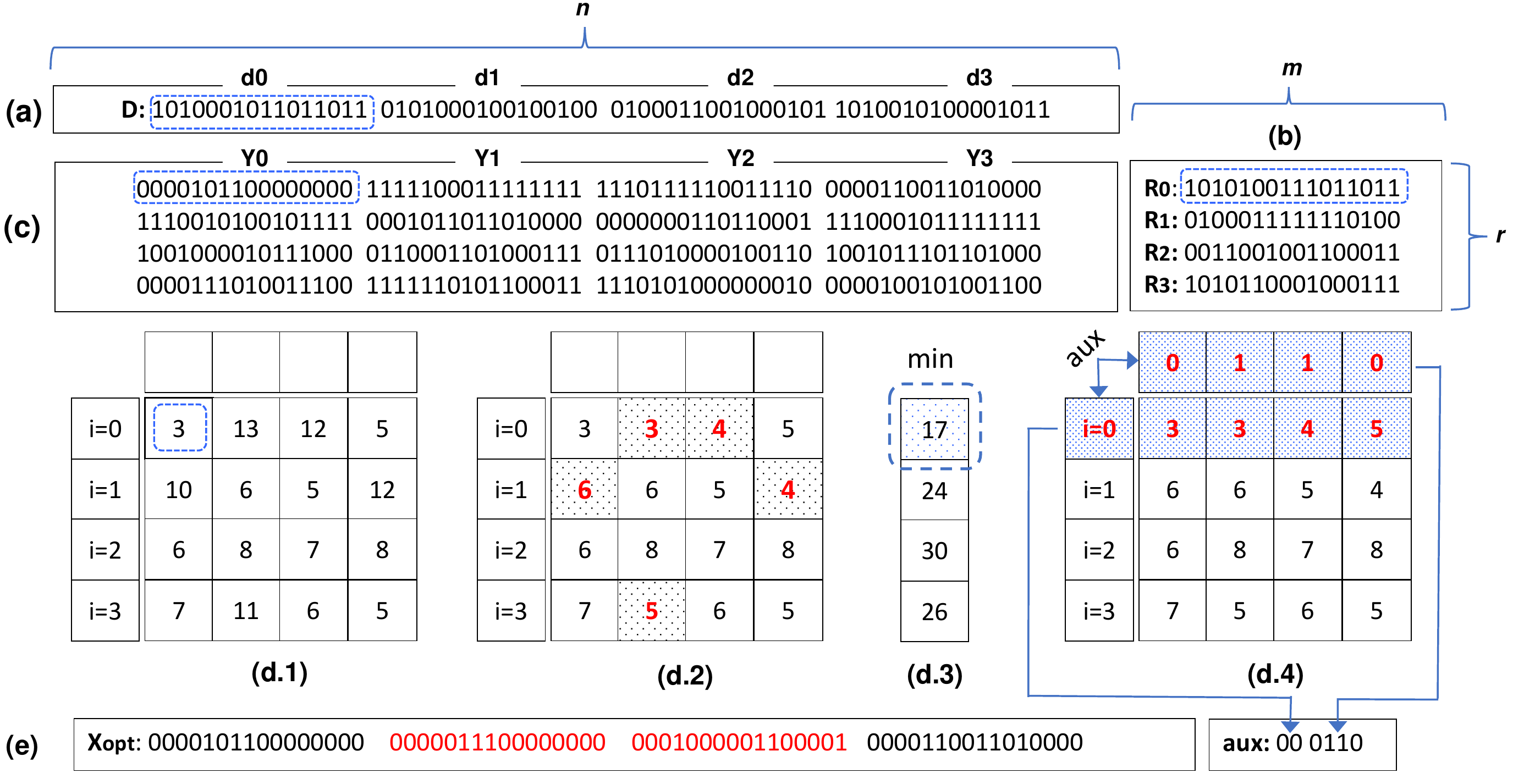}
\vspace{-0.1in}
\caption{Example of VCC for 1's minimization (a) 64-bit encrypted data block; (b) four 16-bit coset kernels; (c) results of coset kernels \texttt{XOR} data sub-blocks; (d.1) cost (number of `1's) for each 16-bit sub-block produced in (c); (d.2) selecting minimum cost of kernel or inverted kernel (for `1's min. the form $<\frac{m=16}{2}$); (d.3) the cost (number of `1's) in the sub-blocks of each row incl. corresponding aux. bits. For example, `17' in the first row of (d.3) is the sum of 3, 3, 4, 5 and 2 (aux. bits); (d.4) the selected row with lowest cost; (e) the output: 64-bit encoded data + 6 aux. bits.
}\label{vcc_example}
\end{center}
\vspace{-0.2in}
\end{figure*}
\section{Virtual Coset Coding (VCC)}
\label{VCC}
\label{sec:VCC}
Virtual Coset Coding is designed to achieve the same benefits as RCC for large numbers of cosets but with reduced encoding complexity.  By relaxing the strict independence of RCC cosets, we can instead compose a virtual coset by concatenating smaller \textbf{coset kernels} in the regular or inverted form.  By applying and evaluating these kernels and their inversion in parallel for kernel sized partitions of the original coset length, we can evaluate many virtual cosets (permutations of the kernels) in short order, reaping the benefits of full-length RCC cosets at a reduced complexity.

Specifically, to generate each $n$-bit random coset candidate, $V$, VCC uses an $m$-bit random string, $R_i$, and concatenates $p=n/m$ copies of this string. Although this reduces the randomness of the bits in $V$, this randomness is restored when $V$ is \texttt{XOR}'d with the $n$-bit random (encrypted) data block $D$, to produce the corresponding code candidate~\cite{katz1996handbook,goldreich2009foundations}.

Furthermore, VCC generates $2^p$ coset candidates from $R_i$ by using either $R_i$ or its complement for each of the $m$-bit partitions of the coset candidate. This sacrifices some independence of the $2^p$ generated candidates but as we will show in Section~\ref{sec:results}, such a loss of independence has a minimal effect on the quality of the encoding.

Finally, we can generate $N$ coset candidates, where $N$ is a multiple of $2^p$, by using $r= \frac{N}{2^p}$ different instances of our random strings, $R_0... R_{r-1}$. We denote the coset encoding scheme that uses $N$ virtual cosets generated from $r$ coset kernels as described above by VCC($n,N,r$).

Thus, instead of applying RCC($n,N$) by \texttt{XOR}ing an $n$-bit encrypted block, with $N$ random coset candidates, VCC($n,N,r$) divides the encrypted data block into $p$, $m$-bit sub-blocks and uses $r$, $m$-bit coset kernels along with the idea of \texttt{XOR}/\texttt{XNOR}ing the sub-block and the coset kernels. 
The \texttt{XOR} and \texttt{XNOR} operations offer two encoding alternatives per sub-block by each coset kernel, resulting in \textit{virtually} increasing the number of encoding alternatives generated by one $m$-bits coset kernel to $2^{p}$, $n$-bit random coset candidates.

To provide a concrete virtual coset coding example, consider applying VCC($n=64$, $N=64$, $r=4$) for a 64-bit data block at an address which for simplicity of illustration contains data consisting of all `0's, noting that normally new data is compared against actual original data read from the memory as shown in Fig.~\ref{fig:vcc_enc_dec_rom}.  In this case, VCC can be applied to reduce the number of bit changes by optimizing for the coset candidate which writes the fewest `1's.  The encrypted input data block $D$ is shown in Fig.~\ref{vcc_example}(a). The first step is to divide $D$ into four ($p$) 16-bit partitions $d_{0},d_{1},d_{2},d_{3}$ as shown in the figure.
To encode $D$, VCC consults the four, 16-bit coset kernels, $R_{0},...,R_{3}$, shown in Fig.~\ref{vcc_example}(b).
Each coset kernel $R_i$, is \texttt{XOR}'d with each of four data sub-blocks to produce the potential encoded sub-blocks $Y_{0},...,Y_{3}$ in Fig.~\ref{vcc_example}(c). 
Then, the number of `1's in each encoded sub-block is counted as depicted in~Fig.~\ref{vcc_example}(d.1).  For instance, the \texttt{XOR} of $d_{0}=``1010001011011011$'' and  $R_{0}=``10101000111011011$'' produces ``$0000\textbf{1}0\textbf{11}00000000$,'' which contains 3 ones, depicted in the left upper entry of Fig.~\ref{vcc_example}(d.1). The other entries in Fig.~\ref{vcc_example}(d.1) are computed in a similar way. In this simple example, we can further improve performance: if the number of `1's in any $d_j$ \texttt{XOR} $R_i$, say $\alpha$, is greater than $\frac{m=16}{2}$, then the number of `1's can be reduced to $m-\alpha$ by replacing $R_i$ with its complement. This is recorded in the array of Fig.~\ref{vcc_example}(d.2), where the entries for which the complement of $R_i$ ($\overline{R_i}$) have been used are highlighted.  For more general cost functions, both $R_i$ and $\overline{R_i}$ must be evaluated.


The next step is to count the total number of `1's in each row of Fig.~\ref{vcc_example}(d.2) as shown in Fig.~\ref{vcc_example}(d.3) and chose the row that minimizes the number of `1's to complete the encoding, as highlighted in Fig.~\ref{vcc_example}(d.4).
In the shown example, the first row ($i=0$), which used the first coset kernel, has the minimum number of `1's, and that minimum is obtained when $d_{0}$ and $d_{3}$ are \texttt{XOR}'d with $R_{0}$ while $d_{1}$ and $d_{2}$ are \texttt{XOR}'d with the complement of $R_{0}$. The final encoded block, $X_{opt}$, is shown in Fig.~\ref{vcc_example}(e), along with six auxiliary bits; two to identify the index, $i$, of the random string used in the encoding and four to identify which of $d_0 ... d_3$ was \texttt{XOR}'d with $R_i$ and which was \texttt{XOR}'d with its complement. These six bits identify the index, $opt$, of the coset candidate used for the encoding. 
Note that each of the four coset kernels $R_0... R_3$ generates 16 virtual coset candidates, for a total of 64 candidates. In the more general case, the optimization could instead capture the number of undesirable symbols, whether that be symbols which are different than the existing data or symbols which are otherwise undesirable, depending on existing symbol in the memory. In this way, the same procedure can be used to optimize for reducing bit changes, matching the value of known faulty cells, which may reduce energy associated with writing certain symbols, or any combination of the above by designing an appropriate cost function.  The resultant cost is encoded in the count, which can then be used to select the best coset as described above.

Note also that, except for the step from Fig.~\ref{vcc_example}(d.1) to Fig.~\ref{vcc_example}(d.2), the computational complexity of exploring the 64 virtual cosets in the above example is the same as exploring four 64-bit cosets if RCC were used. In the more general case where evaluating the kernels or their inverses costs some $\Delta$, for each of the $p$ partitions we apply $r$ coset kernels for a total complexity of $2\cdot \Delta \cdot p\cdot r$.  For RCC with an equivalent count of cosets of length $n$, we are applying $N=r\cdot 2^p$ codes, to $p$ partitions for a total complexity of  $\Delta \cdot p \cdot r \cdot 2^p$.  Using VCC, we improve the computational complexity by a factor of $2^{p-1}$ for which we demonstrate the implementation impact in Section~\ref{sec:hardware}, specifically in Fig.~\ref{fig:HW}.  Thus, the computational complexity of VCC($n,N,r$), where $r=\frac{N}{2^p}$, is similar to RCC($n,r$). Moreover, the look-up table space complexity to store the $r$ random ($\frac{n}{p}$)-bit strings, $R_0, ... , R_{r-1}$ in VCC($n,N,r$) is $\frac{1}{p}$ of the space required to store $r$ random $n$-bit cosets in RCC($n,r$), and the auxiliary information per word is reduced to $log_2(r) + p$ bits to encode the selected kernel and which partitions are inverted.



\subsection{Encoding Algorithm}
\label{vcc_algorithm}

The goal of VCC($n,N,r$) is to find, among the $r$ stored coset kernels, the kernel that should be used in the construction of the optimal virtual coset candidate that minimizes the cost function.
\newlength{\textfloatsepsave} \setlength{\textfloatsepsave}{\textfloatsep} \setlength{\textfloatsep}{0pt}
\setlength{\intextsep}{0pt}
\begin{algorithm}[!b] 
\vspace{-0.1in}
\small
\textbf{Input:}~$D$:~an $n$-bit encrypted data block.\\
\textbf{Output:}~$X_{opt}$:~an optimum candidate; $opt$: a $k=log_{2}(N)$-bit index. \\
\textbf{Function:} $HW(s)$:~This function computes the Hamming weight of the binary string $s$.\\
\textbf{Storage:} $R_{0}...R_{r-1}$: $r$, $m$-bit random coset kernels.\\

Partition $D$ into $m$-bits sub-blocks, $d_{0}...d_{p-1}$\\
\For{$i\leftarrow 0$ \KwTo r-1}{
    \For{$j\leftarrow 0$ \KwTo p-1}{
	    $Y_{j} = d_{j} \oplus R_{i}$;\\
        \If{$HW(Y_{j} < m/2$)}{
            $flag_{j}=0;$ \\
	    }
        \Else{
            $Y_{j} = Y_{j} \oplus ``1 ... 1";$\\
            $flag_{j}=1;$
        }
    }
    $best_i=(i * 2^p) + bin2dec(flag_{0}... flag_{p-1});$\\
    $X_{best_{i}}=Y_{0}...Y_{p-1}$;\\
    $best\_cost_{i} = HW(X_{best_{i}})+HW(best_i)$
}
\If{$ \forall i~~,\exists~i'~~best\_cost_{i'} < best\_cost_i$}{
    Out $opt=best_{i'}$ and $X_{opt}$
}
\vspace{-.05in}
\caption{Virtual Coset Coding~($n,N,r$)}
\vspace{.2in}
\label{vcc_encoding}
\end{algorithm}
\setlength{\textfloatsep}{\textfloatsepsave}
The process of VCC($n,N,r$) encoding is shown in Algorithm~\ref{vcc_encoding}.
First, the input data block, $D$ is partitioned into $m$-bit sub-blocks, $d_{0}...d_{p-1}$. 
Then each of the coset kernels, $R_{0}...R_{r-1}$, are considered independently.
In each iteration of the loop described in lines 6-16, 
$R_i$ is considered and is \texttt{XOR}'d with the sub-blocks $d_{0}...d_{p-1}$ (lines 7-16, can be done in parallel). 
The cost (in this case the number of `1's) of each $d_j \oplus R_i$ is calculated (line~8), and $flag_j$ is used to keep track of any $j$ for which that cost is larger than half of the sub-block size, thus implying the use of the complement of $R_i$ (lines 12-15). When concatenated together, these flags identify which one among the $2^p$ virtual cosets generated by this $R_i$ optimizes the cost.
Line 15 then computes the index of that coset among the $N$ virtual coset candidates. That index is computed from the $\log_2 r$ auxiliary bits identifying $i$ and the $p$ auxiliary bits specified by $flag_0,... flag_{p-1}$.
After identifying the best virtual coset generated by each coset kernel, $R_i$, the optimal virtual coset is identified (in line 18) among the best cosets for $i=0...r-1$. Note that the cost of the auxiliary bits identifying the index of a virtual coset is considered in line 19 when computing the cost of a potential code word.
Finally, the algorithm outputs the optimum virtual coset candidate, $X_{opt}$, and its corresponding index $opt$. In the general case, line 19 evaluates a generic cost function to optimize for specific symbols to reduce stuck-at faults, energy, or programming latency as shown in Fig.~\ref{fig:vcc_enc_dec_rom}. The total search space for finding the optimum candidate in VCC$(n,N,r)$ (Line 7) is scaled down to $r$, which translates to a reduction by a factor of $2^{p}$ over the search space for RCC$(n,N)$ and BCC$(n,N)$.
The process of decoding is simpler: $d_j = (R_{enc} | \overline{R_{enc}}) \oplus Y_j$ based on the encoded value $enc$ and $flag_j$. This only requires \texttt{XOR} or \texttt{XNOR} to retrieve $D$ and incurs negligible latency overhead. 
Finally, we note that the number of auxiliary bits for every $n$ bits of data 
in VCC($n,N,r$) is $\log_2 N$, which is the same number of auxiliary bits as in RCC($n,N$). Specifically, it uses $\log_2 r$ auxiliary bits for identifying the coset kernel + $\log_2 \frac{N}{r}$ auxiliary bits for keep tracking of the flag bits. For example, given $r=16$ coset kernels, $N=256$ independent random coset candidates and $n=64$ bits block size, both $RCC(64,256)$ and $VCC(64,256,16)$ consume 8 auxiliary bits, matching the 12.5\% capacity supported for memories employing SECDED.

\subsection{Coset Vector Generator for MLC PCM}
 Table~\ref{tab:symbol_transition} shows symbol transitions from old to new states for MLC PCM, based on the prototype device in~\cite{MLC_Energy}, where Gray coding is used to encode the sequence of states spanning the range of resistance levels in each cell. An important observation from the table is that a high energy transition happens when the right digit of a new symbol is 1.  Any encoding can focus on modifying these digits to achieve the most benefit for write energy.  Furthermore, since the write energy is insensitive to the left digit, we can leave it unchanged, and use its encrypted value to generate cosets at runtime. This removes a potential vulnerability that the kernels may be discovered and used to defeat the scheme for reducing energy and improving endurance.  
 
Since the write energy is insensitive to the value of the left digit of each symbol, and we have effectively random data written back due to encryption, we can apply encoding only to the right digit, and use the left as a random seed to generate coset kernels.  Algorithm~\ref{vcc:base_vector_generator} is an example of one such generator. To generate $r$ $m$-bit random coset kernels, $l=\frac{n}{2}$ left binary digits are extracted from $\frac{n}{2}$ symbols of the encrypted data block~(line 1). Then, these binary digits are divided into $b=\frac{l}{m}$ base vectors where $\frac{r}{b}$ vectors are derived from each base vector. Specifically, the generator uses $1+log_{2}\frac{r}{b}$-bit unique masks\footnote{One extra bit in the mask allows to remove complementary patterns from the random binary vectors.} (line 3) that are  \texttt{XOR}ed with each base vector. More specifically, each base vector is divided into $1+log_{2}\frac{r}{b}$-bit sub-vectors and each mask is independently \texttt{XOR}ed with sub-vectors (lines 5-8). For example, consider Fig.~\ref{vcc_example}(a) where 32 left digits are divided into two base vectors `1101101100000100' and `0001000011000011'. Given $r=4, m=16, n=64$ and $b=2$, two 2-bit masks are generated $M_{0}=00$ and $M_{1}=01$. These masks are \texttt{XOR}ed with two base vectors to generate 4 random binary vectors as `1101101100000100', `1000111001010001', `0001000011000011' and `0100010110010110'.

\begin{table}[!t]
\centering
\caption{Symbol energy transitions from Old (O) to New (N) states.  Energy based on prototype MLC PCM~\cite{MLC_Energy}.}
\label{tab:symbol_transition}
\begin{tabular}{l|c|c|c|l|}
\cline{2-5}
                         & \multicolumn{1}{l|}{N(00)} & \multicolumn{1}{l|}{N(01)} & \multicolumn{1}{l|}{N(11)} & N(10)                     \\ \hline
\multicolumn{1}{|l|}{O(00)} & -                       & high                       & high                       & low                      \\ \hline
\multicolumn{1}{|l|}{O(01)} & low                       & -                       & high                       & low                      \\ \hline
\multicolumn{1}{|l|}{O(11)} & low                       & high                       & -                       & low                      \\ \hline
\multicolumn{1}{|l|}{O(10)} & low                       & high                       & high                       & \multicolumn{1}{c|}{-} \\ \hline
\end{tabular}
\vspace{-0.2in}
\end{table}

\setlength{\textfloatsep}{0pt}
\setlength{\intextsep}{0pt}
\begin{algorithm}[!b]
\small
\textbf{Input:}~$L$:~ $l=\frac{n}{2}$ left binary digits extracted from the encrypted data block that are divided into $b=\frac{l}{m}$ base vectors.\\
\textbf{Output:}~$R_{0},...,R_{r-1}$: $r$ random coset kernels of length $m$. 
\\
\textbf{Mask:} $M_{0},...,M_{\frac{r}{b}-1}$:  $\frac{r}{b}$ binary sequences of length $1+log_{2}\frac{r}{b}$.\\

\For{$i\leftarrow 0$ \KwTo $\frac{r}{b}$-1}{
\For{$j\leftarrow 0$ \KwTo b-1}{
    $M_{i} = dec2bin(i)$\\

    $R_{i\times b+j} = M_{i} \oplus L\left[j\cdot(\frac{l}{b}):(j+1)\cdot(\frac{l}{b})-1\right]$
}}
\caption{Base Coset Vector Generator}\label{vcc:base_vector_generator}
\label{algo:generated-cosets}
\end{algorithm}
\setlength{\textfloatsep}{\textfloatsepsave}

\subsection{On-chip Encoding/Decoding Architecture}\label{vcc_architecture}
\begin{figure}[!t]
\begin{center}
\includegraphics[width=0.475\textwidth]{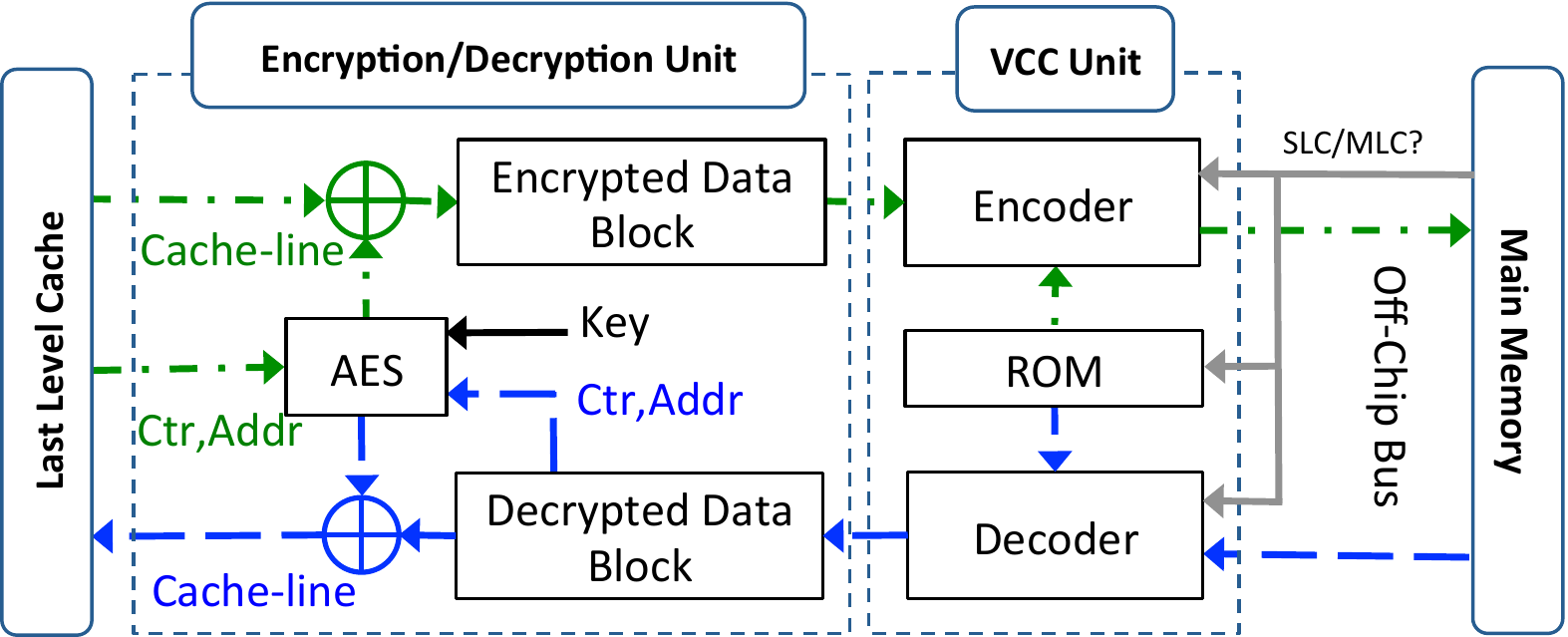}
\vspace{-0.1in}
\caption{On-chip integration of VCC with the encryption/decryption units.}\label{cache_I/O_memory_structure}
\end{center}
\vspace{-0.2in}
\end{figure}

The main architecture of our scheme is illustrated in Fig.~\ref{cache_I/O_memory_structure} where, before being written to memory
a 512-bit cache-line is encrypted through an \texttt{XOR} operation with 4$\times$128-bit \textit{random binary streams}, sometimes called a one-time pad, generated by the AES engines. 
These streams are unique as the corresponding counter of the cache-line, the cache-line address, and the 256-bit unique key are fed to the AES engines to generate these random binary streams. For every write request from last-level cache (LLC) to the main memory, the four AES engines increment the value of the cache-line counter by 1 in order to generate unique streams for each stored value. The new counter value is stored along with the cache-line to allow decryption when reading the line. The encrypted line is then forwarded to the VCC encoder which partitions it into 8, 64-bit blocks to be encoded. The encoded blocks along with the corresponding auxiliary bits are transmitted on the bus. If the coset kernels are to be generated and stored in advance, they are placed in the optional ROM for use during the encode and decode processes.  Otherwise, the ROM is omitted, and the kernels are generated using Algorithm~\ref{vcc:base_vector_generator}. The decoding is the inverse of the encoding process. Specifically, the received data blocks from the off-chip bus are decoded and then decrypted before being written in the LLC.
Note that both the decoding and decryption processes use simple \texttt{XOR} operations.

\begin{figure}[!t]
\begin{center}
\includegraphics[width=0.475\textwidth]{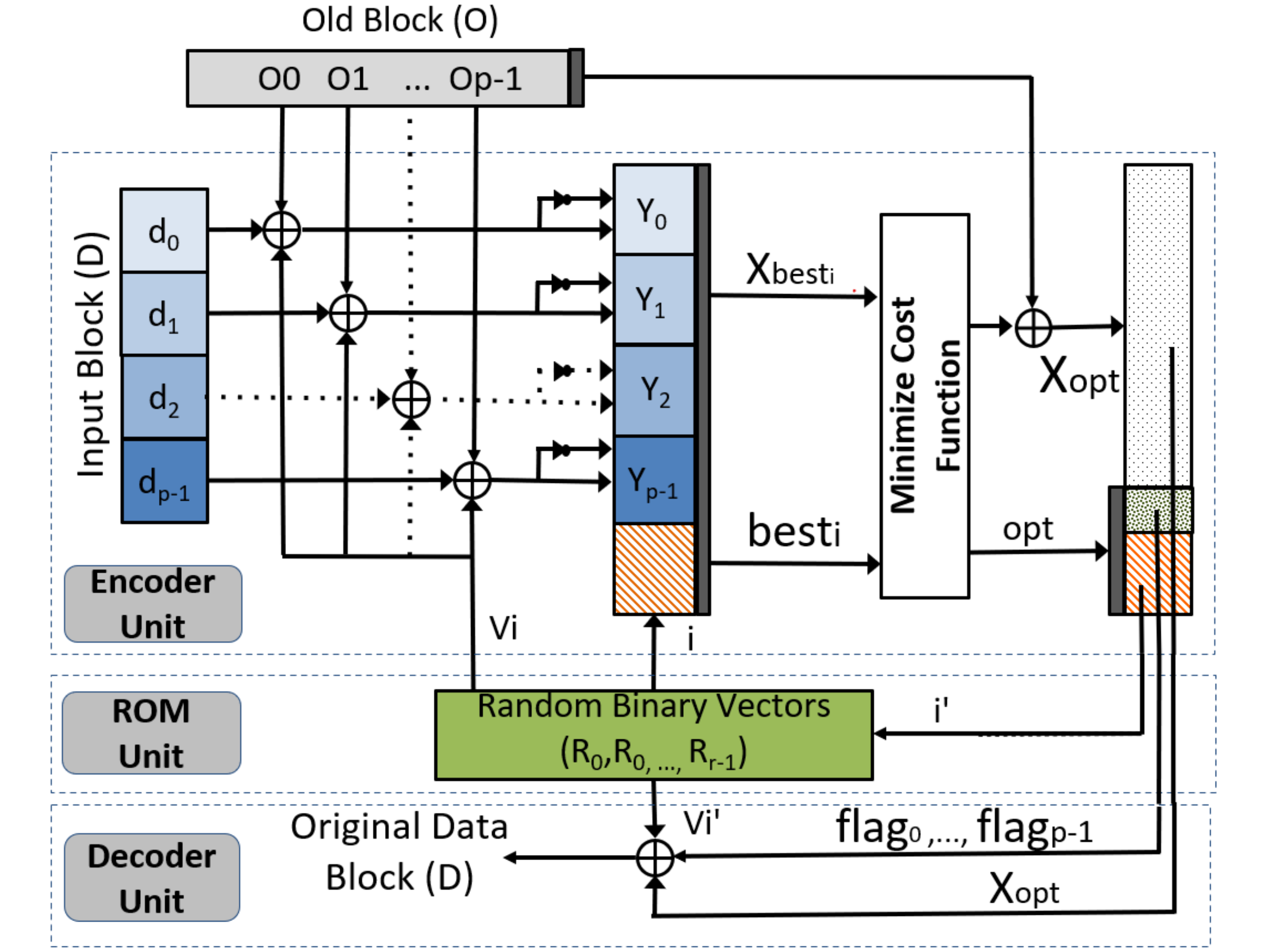}
\vspace{-0.1in}
\caption{The VCC architecture including the encoder unit, the optional ROM unit, and the decoder unit.}\label{vcc_enc_dec_rom}\label{fig:vcc_enc_dec_rom}
\end{center}
\vspace{-0.2in}
\end{figure}

\begin{figure*}[tbp]
\subfigure[Area]{
    \includegraphics[width=.265\pdfpagewidth]{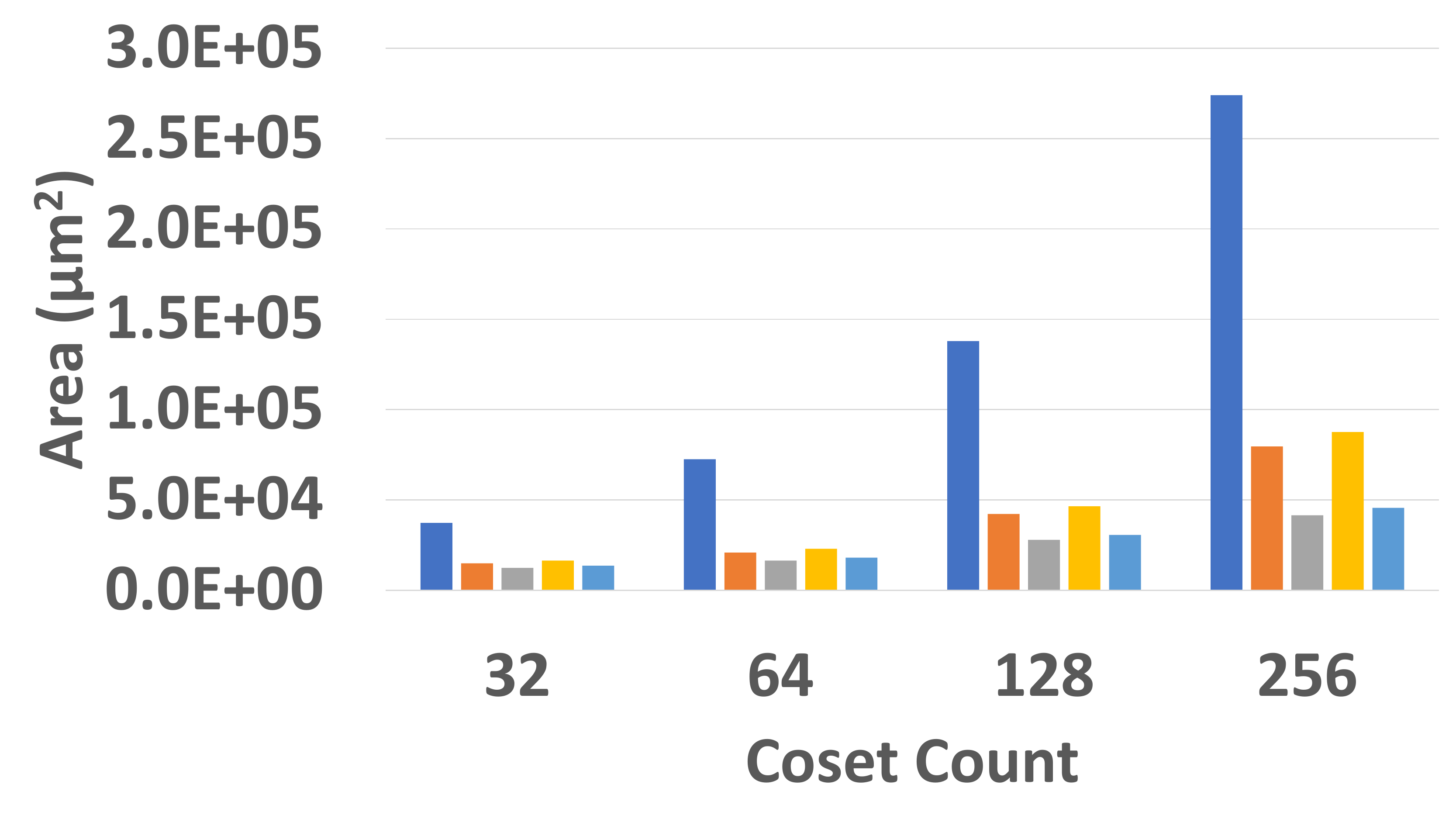}
    \label{fig:areaHW}
}
\subfigure[Energy]{
    \includegraphics[width=.265\pdfpagewidth]{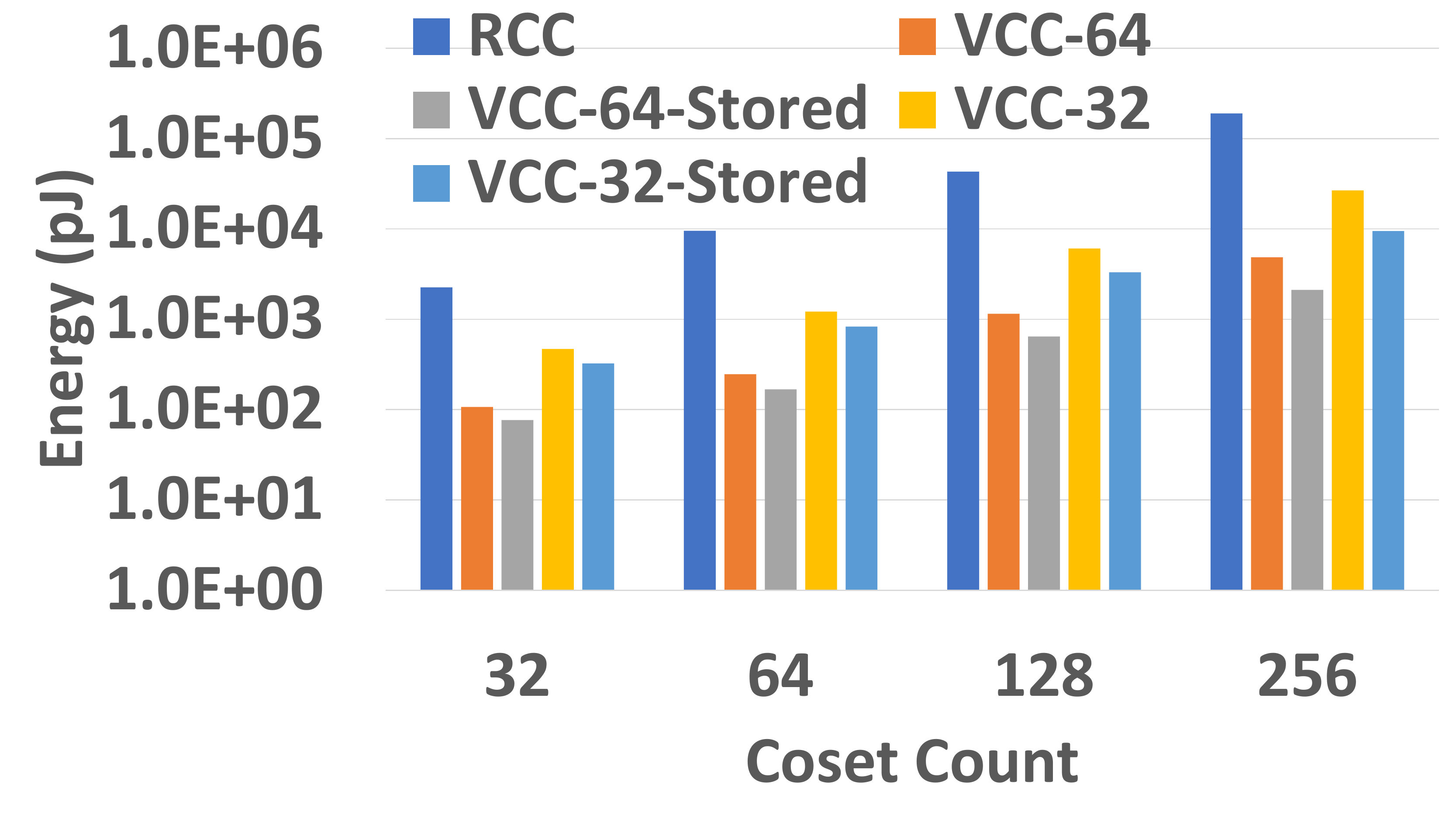}
    \label{fig:energyHW}
}
\subfigure[Delay]{
    \includegraphics[width=.265\pdfpagewidth]{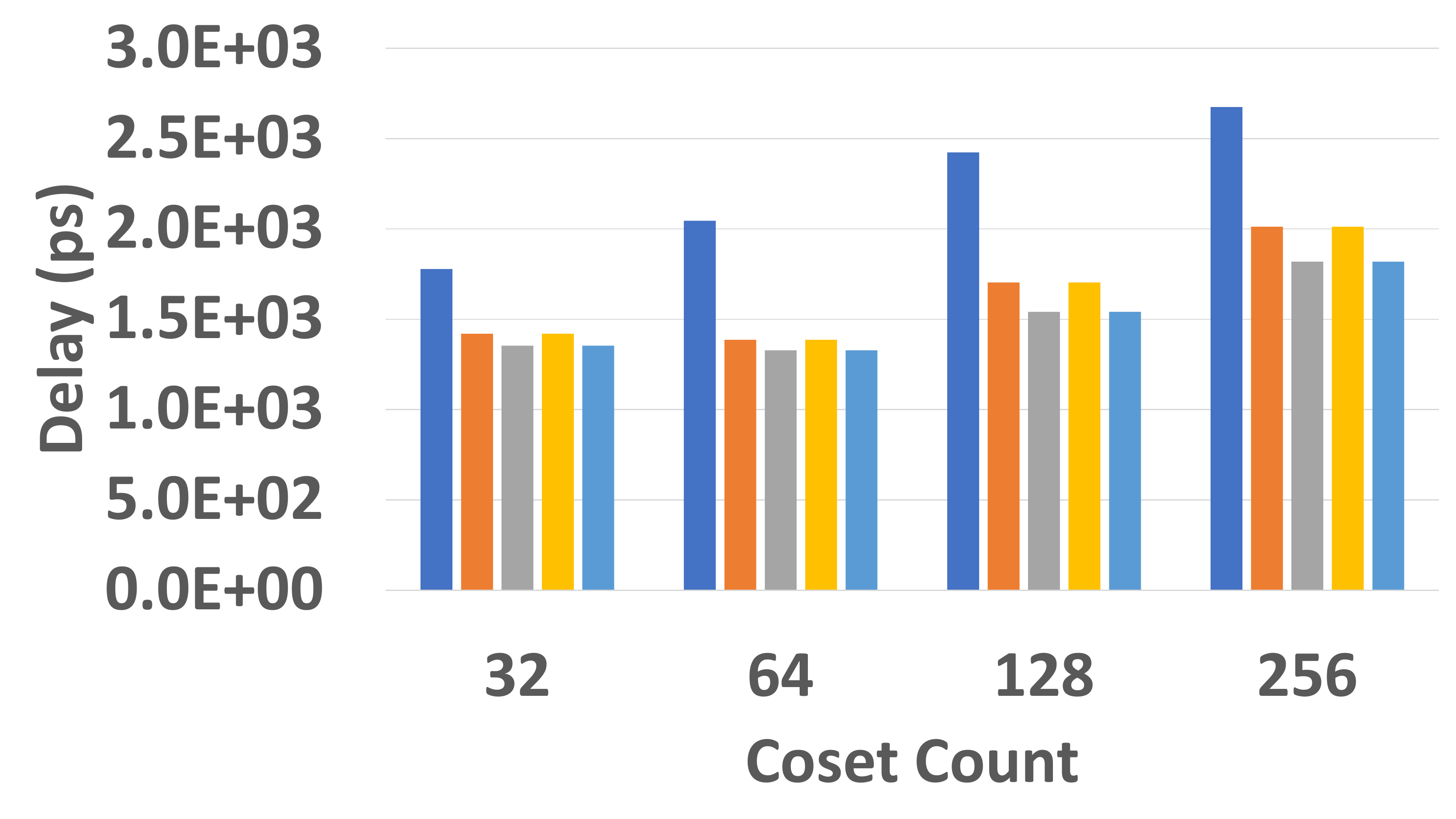}
    \label{fig:delayHW}
}
\vspace{-0.1in}
\caption{Coset encode hardware in 45nm technology.  For VCC, the coset count represents equivalent number of cosets.}
\label{fig:HW}
\vspace{-0.2in}
\end{figure*}

We depict the on-chip VCC architecture in more detail in Fig.~\ref{vcc_enc_dec_rom} for a 64-bit encrypted block, $D$. First, $D$ is partitioned into $p$, $m$-bit encrypted data sub-blocks~$d_{0}...d_{p-1}$. The random strings, $R_{0}...R_{r-1}$, stored in a ROM, are fetched and each string $R_{i}$, $0\leq i \leq r-1$, is \texttt{XOR}'d and \texttt{XNOR}'d with $d_{0},d_{1}...d_{p-1}$ in parallel and the flags $flag_0... flag_{p-1}$ are set depending on which operation, \texttt{XOR} or \texttt{XNOR}, minimizes the defined objective. In the case of generated cosets, the ROM Unit is instead replaced by a Coset Generator Unit, which provides random strings generated from the encrypted data block $D$ per Algorithm~\ref{vcc:base_vector_generator}. The value of $i$ and the flags are used to form the index, $best_i$, and the output with the minimum cost is stored in $X_{best_i} = Y_{0}... Y_{p-1}$. Finally, the optimum candidate $X_{opt}$ is selected from among $X_{best_0}... X_{best_{r-1}}$ and stored into the memory. The decoding requires fewer logic operations; 
the index $i'$  is fed to the ROM to retrieve the coset kernel $R_{i'}$, which is either \texttt{XOR}'d or \texttt{XNOR}'d with each of the $p$, $m$-bit partitions of the received data block, $X_{opt}$, depending on  $flag_{0}... flag_{p-1}$, to retrieve the original encrypted block.


\section{Design Space Exploration}
\label{sec:design-space}\vspace{-0.03in}
There are many configurations of VCC which may be applied to a given data block of size $n$, including the granularity at which we apply VCC, the number of partitions within that granularity, and the count of coset kernels to be applied.  For our evaluations, we held the overall data block size fixed to the common 64-bit word size, as ECC (64,72) is already configured to protect this word size and establishes an upper bound for the amount of auxiliary bits allowed for each data block.  Namely, we have at most eight bits per 64-bit word, or 64 bits per 512-bit cache line being written to memory, with which to encode auxiliary information.  We also explored a 32-bit data block size to correspond with legacy machines.  We evaluated hardware implementations of VCC and RCC and used this to inform design space explorations of energy savings.  We also considered reliability in the form of protection against writing a value into a stuck-at cell with a different value, noted as stuck-at wrong (SAW) bits.  This allows the identification of appropriate VCC configurations to select for deeper analysis.

\subsection{Hardware Evaluation}
\label{sec:hardware}
To explore the impact of parameter choice on the hardware design of a VCC unit, we implemented a VCC  encoding unit in a hardware description language and synthesized the implementation to an ASIC design using the Cadence\textsuperscript{\textregistered} 
Encounter RTL Ultra synthesis tool suite~\cite{cadence_encounter} targeting 45nm process technology.  We implemented every possible permutation for $n\in\{32,64\}$ with $N\in\{32,64,128,256\}$. Our preliminary results showed little difference between $m=16$ and $m=32$, thus we report results for $m=16$ as in our example. 
We implemented two versions of each design, one which generates coset kernels via Algorithm~\ref{vcc:base_vector_generator}, and one which stores pre-generated random coset kernels in a ROM Unit.  These are labeled as VCC[32/64] and VCC[32/64]-Stored, respectively, where 32/64 refers to the value of $n$.  For comparison, we also implemented RCC(64,$N$) over the same granularity, where $N$ random cosets of length 64 are stored in a ROM.  The decoder circuit is of negligible complexity and overhead compared to the encoder circuit, even for generated cosets, because only the stored coset mask must be recovered.  Thus, we report results for the block of interest, the encoder.



Fig.~\ref{fig:areaHW} shows the overall on-die area across the designs under test in $\mu m^2$.  Our design evaluated each partition and coset kernel in parallel targeting a delay-optimized design.  
Area increases slightly for all VCC designs for larger coset counts with a slightly sharper trend for generated cosets.  RCC has a much higher starting area and increases at a substantially faster rate.  The additional area alone demonstrates RCC's impracticality in terms of recent overhead metrics~\cite{VOGEL_DRAM_ENERGY,HUANG_DRAM_SCALING}.
Fig.~\ref{fig:energyHW} depicts the operating energy across the designs. 
RCC energy starts an order of magnitude higher than VCC and that gap increases with coset count.  VCC-32 energy is monotonically larger than VCC-64. 
%
%
Fig.~\ref{fig:delayHW} provides the encoding delay when writing new data. 
The difference in latency follows similar trends where VCC holds its latency to 1.8-2ns at 256 cosets and RCC exceeds 2.6ns for 256 cosets, making it a higher performance overhead.



Thus, we can conclude that VCC outperforms RCC, is best applied at the 64-bit granularity, and stored vs. generated cosets are equally if not more practical in the system.  
In the next section, we evaluate the effectiveness of VCC in energy and reliability improvement against the baseline of RCC.

\subsection{VCC Coset Effectiveness \& Scalability}

Given VCC is an approximation of RCC we 
%
%
conducted a preliminary energy study with randomly generated data written to a small memory 100,000 times to compare RCC, with VCC and VCC-stored.  
To evaluate the utility of an increased number of cosets, we compared VCC(64,32,2), VCC(64,64,4), VCC(64,128,8), and VCC(64,256,16), using the MLC PCM symbol energy reported in the literature~\cite{MLC_Energy}. 
\begin{figure}[!b]
\vspace{-.3in}
\begin{center}
\includegraphics[width=0.44\textwidth,trim={0 1.2in 0 0 1.4in},clip]{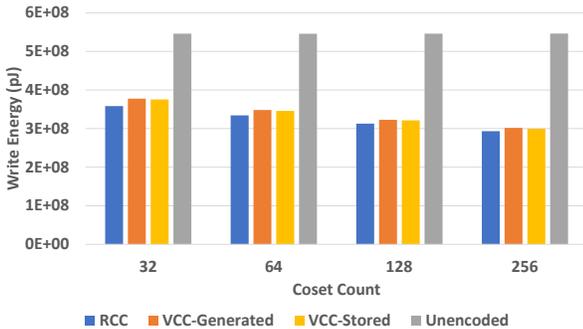}
\vspace{-0.1in}
\caption{Write energy comparison of VCC, VCC using stored cosets, RCC, and unencoded writeback.  Includes the cost of writing auxiliary information.}
\label{fig:rcc_vs_vcc}
\end{center}
\end{figure}

Fig.{~\ref{fig:rcc_vs_vcc}} shows a comparison of write energy of VCC and RCC compared to an unencoded baseline.
RCC reduces write energy about 46.3\% in the 256 cosets case. 
VCC with generated cosets provides about 44.8\% energy savings along with its security advantage, while stored cosets only improves this to about 45.1\% energy savings.  Furthermore, as the number of cosets increases the difference between RCC and VCC decreases.  Thus, both VCC techniques can approximate the best possible energy savings with RCC without the dramatic hardware overhead discussed in Section{~\ref{sec:hardware}}. 
From here on we assume VCC uses generated cosets unless stated otherwise.  



\begin{figure}[tp]
\begin{center}
\includegraphics[width=0.475\textwidth,trim={0 1.2in 0 0.9in},clip]{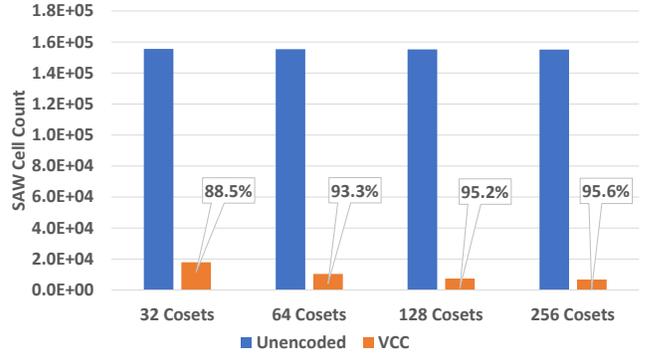}
\vspace{-0.1in}
\caption{SAW cell improvement over unencoded writeback as a function of coset cardinality for a fixed fault incidence rate with MLC PCM. }\label{fig:coset_SAW}
\end{center}
\vspace{-0.35in}
\end{figure}

Similarly, as discussed in Section~\ref{sec:motivation}, more coset candidates can provide a better chance to match SAW cell values (see Fig.~\ref{fig:coset_sweep}). Fig.~\ref{fig:coset_SAW} depicts the results of using VCC to reduce SAW cell values in the presence of failed cells for a fixed fault incidence rate of $10^{-2}$.  This represents a ``snapshot'' of a memory with extreme wear to stress-test VCC.  The reported values are an average across all benchmarks under test, and across five test iterations.  We observe that even a small number of cosets achieves an approximately 88.5\% reduction in SAW values but increasing the number of cosets to 256 can reach more than 95\% reduction. 
%
%
While at this fault rate VCC alone cannot achieve fault tolerance, it can be combined with a simple fault tolerance such as SECDED to be effective for a much higher fault rate than SECDED alone, increasing effective lifetime. This demonstrates the benefit of larger numbers of cosets generated by VCC provides a benefit to energy savings and reliability.  From this point we will show VCC results for 256 cosets unless specified otherwise.

\vspace{-.035in}
\section{Results}
\vspace{-.035in}
\label{sec:results}

In this section we present and discuss the results of our evaluation of energy reduction followed by fault mitigation, lifetime improvement, and performance based on the design space exploration conducted in the previous section.

\vspace{-.035in}
\subsection{Experimental Setup}
\vspace{-.035in}
To evaluate dynamic energy and SAW cell improvement under VCC for MLC PCM we simulated by writing to a MLC memory device with an extreme fault rate, applying VCC based on the design candidates determined from the previous section using $n=64$ and $m=16$.  Fault maps were pre-generated to match a 2GB memory with a fixed $10^{-2}$ fault incidence rate (no wear accumulated during the test). Results presented represent the average of execution with five distinct $10^{-2}$ fault map permutations.  Simulation trace data was generated using a representative subset~\cite{SPEC2017_Survey} of the most memory-intensive benchmarks in the ``SPECspeed\textsuperscript{\textregistered} 2017 Integer'' and ``SPECspeed\textsuperscript{\textregistered} 2017 Floating Point'' categories of the SPEC\textsuperscript{\textregistered} \footnote{SPEC\textsuperscript{\textregistered}, SPEC CPU\textsuperscript{\textregistered}, and SPECspeed\textsuperscript{\textregistered} are registered trademarks of the Standard Performance Evaluation Corporation}
2017 benchmark suite~\cite{SPEC2017}, capturing the raw data and address of evicted data being written back from the last level cache to memory.  Each memory address was initialized with pseudo-random values output from a cryptographically strong byte generator from the OpenSSL software library~\cite{OpenSSL}.  The same generator was used to generate a one-time-pad for counter-mode encryption~\cite{NIST-counter-mode} of the trace data, representing the data generated by the Encryption/Decryption unit of Fig.~\ref{cache_I/O_memory_structure}.
Tolerating any spatial overhead and accounting for the energy required to write the appropriate auxiliary bits, we varied the number of coset kernels $N\in\{32,64,128,256\}$, for VCC using configurations VCC(16,$N$,$\frac{N}{16}$), measuring the write energy associated with the selected coset.  As described in the previous section, cosets were generated using Algorithm~\ref{vcc:base_vector_generator} for VCC on MLC PCM.  Two cost functions were explored, one for first minimizing energy, and then minimizing SAW cells and one for first minimizing SAW cells and then minimizing energy.  We report SAW minimization to show the potential improvement in SAW errors as a metric, not to guarantee correctness.  We note for lower fault rates, it may be possible for VCC to eliminate all SAW errors.

To evaluate VCC lifetime improvement, we simulated a 2GB memory in a system with 4kB pages consistent with previous studies~\cite{AEGIS,retrofit}, using a cache line size of 512 bits.  For each cell therein, a lifetime was assigned according to a normal distribution about a mean of $10^8$ writes~\cite{xue2011emerging}, varied on a normal distribution using a coefficient of variation of 0.2~\cite{retrofit}.  After any given cell exceeded its lifetime, its value was held fixed for the duration of the simulation to represent a ``stuck-at'' cell. This is distinct from the other experiments where fault rate was held fixed; each cell is functional until it exceeds its nominal lifetime.  Each cell was initialized with a pseudo-random value by the same method described above and fed simulation trace data as described above.  Each trace was repeated until four row addresses experienced uncorrectable faults~\cite{start-gap}, and the presented results are an average of 5 distinct lifetime experiments.  We evaluated VCC using the parameters described in Section~\ref{sec:design-space}. 
The cost function first selected SAW minimization followed by energy minimization. In these experiments, the memory fails when SAW errors cannot be entirely eliminated by VCC.  
Write energies for each symbol were computed using the device described in~\cite{MLC_DEVICE}.  For comparison, we simulated how iso-area configurations of SECDED ECC, ECP3, DBI, Flipcy, and RCC perform, with all coset techniques following the same optimization scheme.

\setlength\tabcolsep{3pt}
\begin{table}[t]

\centering
\caption{Architecture parameters for performance study. 
}
\begin{tabular}{c|c|c|c|c|c|c|c}
\hline
\multicolumn{4}{c|}{\textbf{CPU}} & \multicolumn{4}{c}{\textbf{Cache}}\\\hline
\multicolumn{4}{c|}{4 out-of-order cores} & \multicolumn{4}{c}{Private L1 32kiB Inst, 32kiB Data}\\
\multicolumn{4}{c|}{4 issue width} & \multicolumn{4}{c}{Private L2 256kiB/Core}\\
\multicolumn{4}{c|}{28 nm Technology} & \multicolumn{4}{c}{Associativity: 8 (L1 data and L2)}\\
\multicolumn{4}{c|}{1GHz Frequency} & \multicolumn{4}{c}{Block Size: 64B }\\
\hline
\multicolumn{8}{c}{\textbf{Memory: PCM }}\\\hline
\multicolumn{8}{c}{512 bit rows, 64 bit words, 2GiB main memory}\\
\multicolumn{8}{c}{2 channels,
1 rank per channel, 8 banks per rank}\\
\multicolumn{8}{c}{Baseline access delay 84ns~\cite{sniper_new}}\\

\hline

\end{tabular}
\label{tab:arch_parameters}
\vspace{-0.2in}
\end{table}
\setlength\tabcolsep{6pt}
To assess the impact of our techniques on performance, we used the SNIPER full-system simulator~\cite{sniper_new} using parameters shown in Table~\ref{tab:arch_parameters} for each benchmark under test.  Dirty evictions are sent to the encryption unit in parallel with a read request to retrieve the original data.   Writes only commit after this delay and encoding delays (\textit{i.e.,} Read-Modify-Write) reported in our hardware implementation in Fig.~\ref{fig:delayHW}.  


\subsection{Energy Evaluation}
\label{sec:energy_evaluation}
Fig.{~\ref{fig:coset_energy_benchmark}} shows the results of our energy savings with VCC when applied to energy and reliability optimization.  Here, we show two runs, the first of which optimizes for reduced energy first and SAW fault mitigation second (Opt. Energy) and the second \textit{vice-versa} (Opt. SAW).
When optimizing energy we can still minimize SAWs in the presence of endurance faults among the lowest energy encoded candidates.  What is interesting is that the $\approx 28$\% average energy savings achieved are maintained even if we change the cost function to minimize SAWs first and then minimize energy among the candidates with the best reliability.  In both cases, the presented results are for VCC(64,256,16) using MLC PCM.  
Note that energy associated with running the encoder/decoder hardware (see Fig.~\ref{fig:energyHW}) is many orders of magnitude lower than the energy saved; this demonstrates that VCC can significantly reduce dynamic energy over the memory lifetime.

\subsection{Reliability Evaluation}
\label{sec:lifetime_evaluation}

\begin{figure}[tp]
\begin{center}
\includegraphics[width=0.498\textwidth]{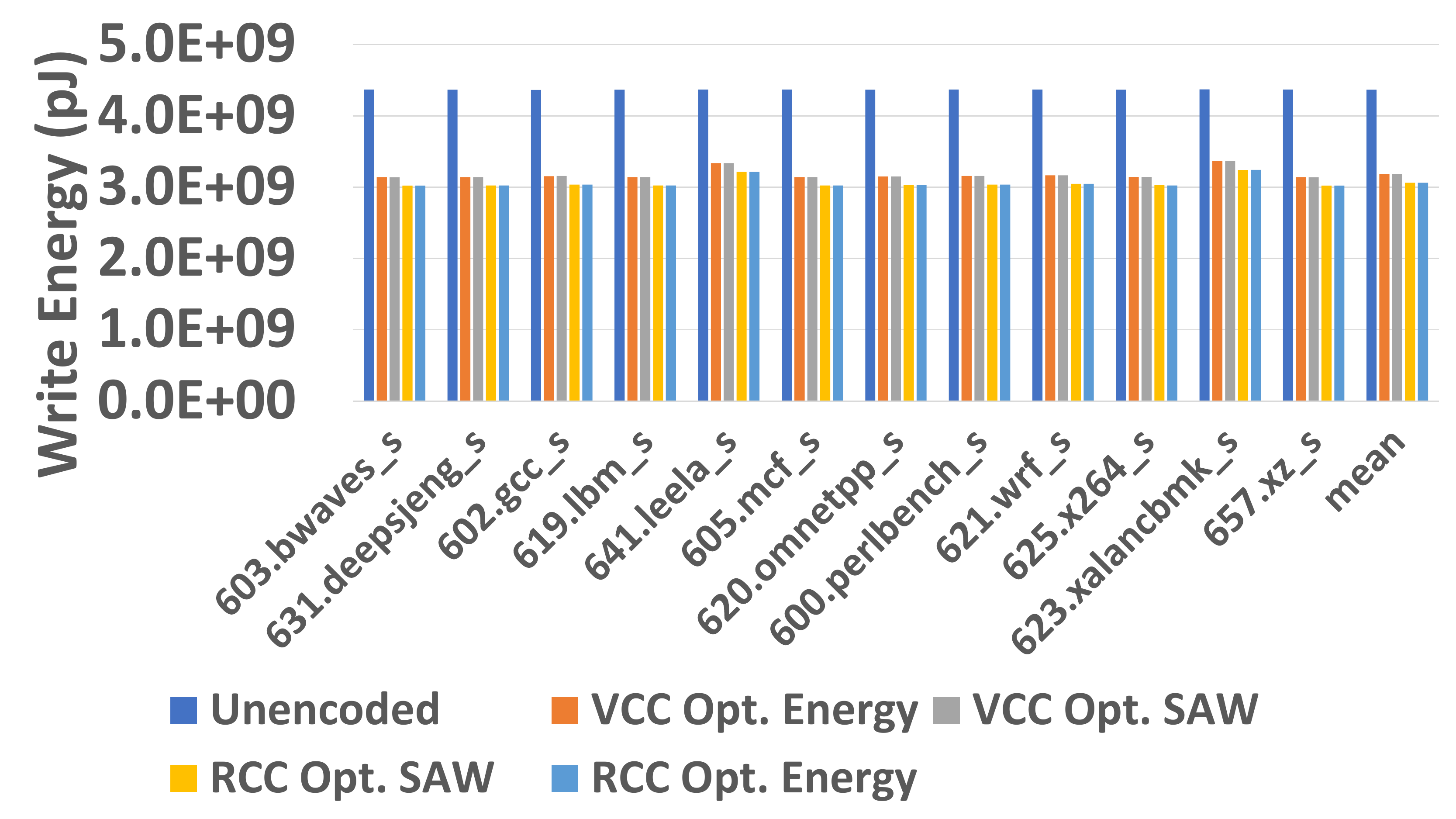}
\vspace{-0.3in}
\caption{Energy improvement across benchmarks over unencoded writeback for a fixed fault incidence rate with MLC PCM.  The cost function optimized for fault mitigation first, and energy second.  Energy calculation includes the cost of writing auxiliary information.}\label{fig:coset_energy_benchmark}
\end{center}
\vspace{-0.2in}
\end{figure}

\begin{figure}[tp]
\begin{center}
\includegraphics[width=0.5\textwidth]{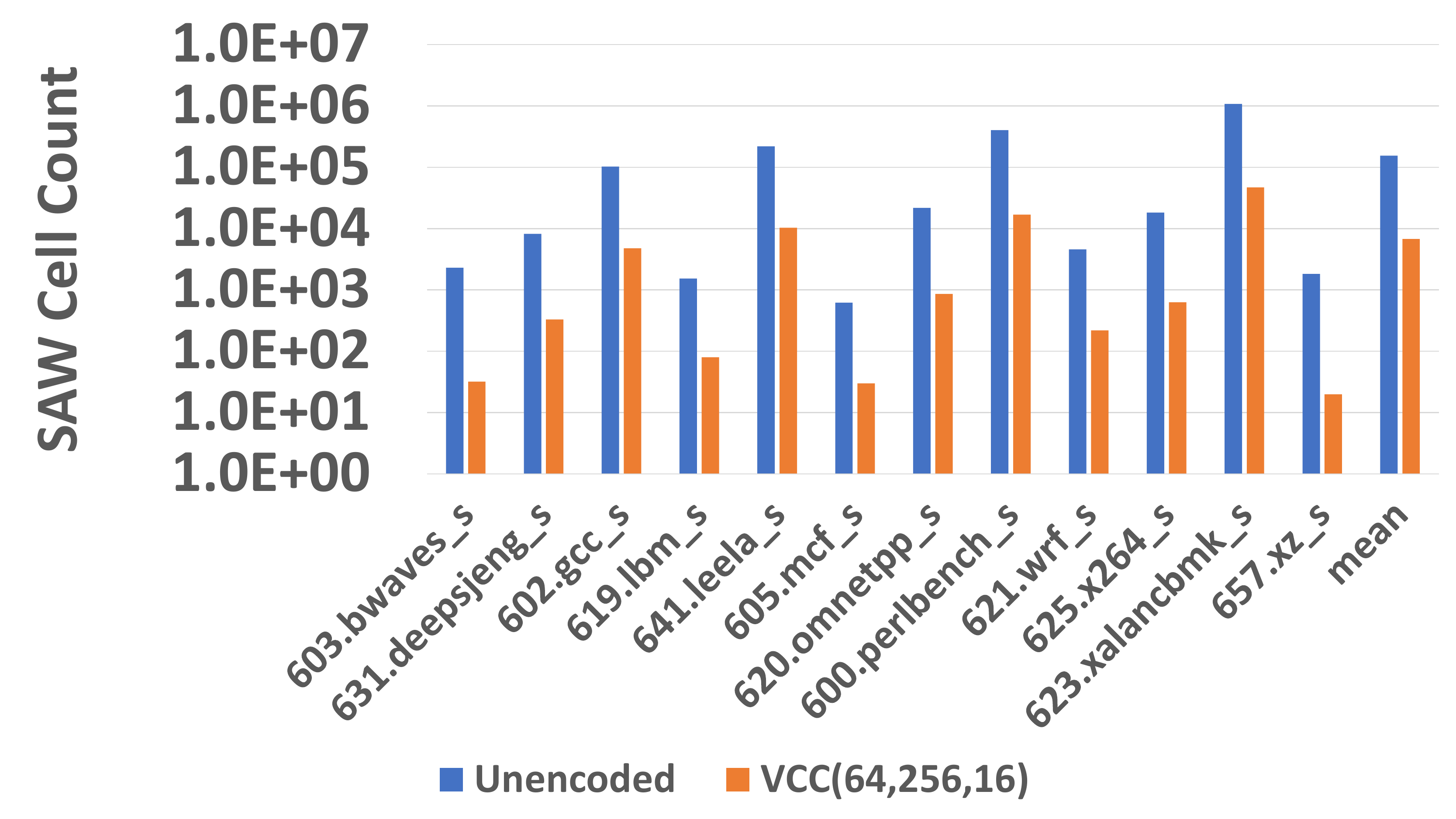}
\vspace{-0.2in}
\caption{SAW cell improvement over unencoded writeback for fixed fault incidence rate on MLC PCM compared to VCC with 256 virtual cosets.}\label{fig:coset_SAW_benchmark}
\end{center}
\vspace{-0.3in}
\end{figure}

Fig.~\ref{fig:coset_SAW_benchmark} shows the reduction of SAW cells across benchmarks, using a pre-generated fault map with a fixed fault incidence rate of $10^{-2}$.  This data represents the same test run as in Fig.~\ref{fig:coset_energy_benchmark}, where the cost function optimizes for SAW mitigation first and energy second.  Here, we can see that the performance of the cosets varies across benchmarks, but still reduces the overall count of SAW cells by at least 95\%.  This result demonstrates the potential of VCC (and cosets in general) to extend the lifetime of a limited-endurance memory life PCM.  
However, for correct execution, 95\% correction will be inadequate.  In order to judge the practical impact of the SAW reduction, it is critical to look at improvement in lifetime as endurance faults accrue.

\begin{figure*}[!ht]
\begin{center}
\includegraphics[width=0.9\textwidth]{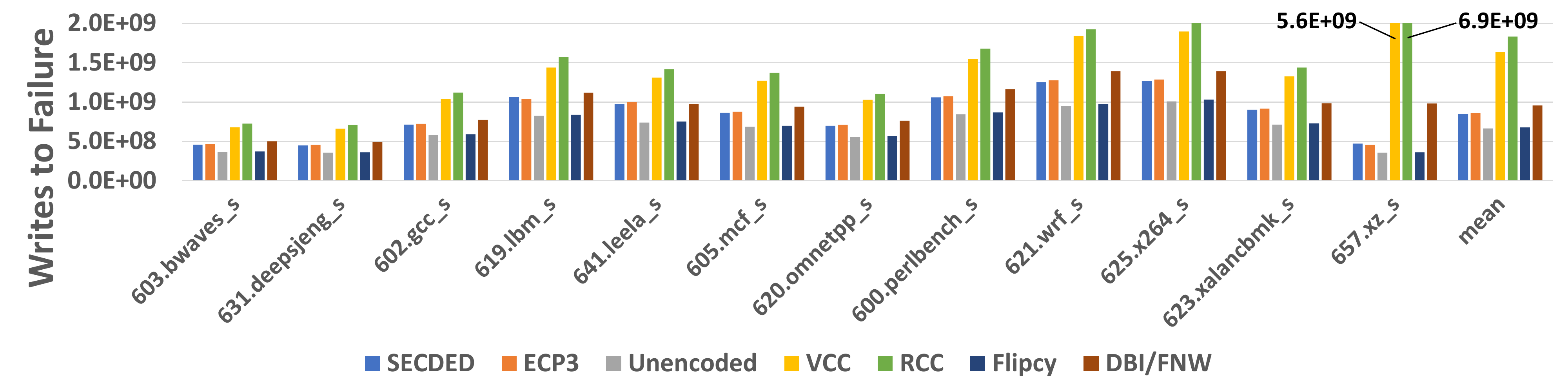}
\vspace{-0.1in}
\caption{Comparison of lifetime writes among protection techniques, applied at a word size of 64 bits to a 2 GB memory.  RCC and VCC use 256 cosets.
}\label{fig:lifetimeComparison}
\end{center}
\vspace{-0.35in}
\end{figure*}
Fig.~\ref{fig:lifetimeComparison} shows the per-benchmark lifetime of each technique for 256 cosets, where the memory lifetime is defined as the count of row writes before four rows fail to write correctly due to SAWs.  
All of the benchmarks under test were selected because they exhibited a large proportion of store instructions, and the distribution of those writes, filtered through the cache, affects how wear is distributed across the cells in the underlying memory depending on the distribution of weaker cells to endurance~\cite{retrofit}.  
Flipcy generally performs poorly due to its granularity and sensitivity to patterns in the data being written.  SECDED, ECP, and DBI/FNW provide a 10-20\% improvement in most cases.  VCC more than triples the lifetime over unprotected and more than doubles the lifetime over SECDED, ECP, or DBI/FNW, nearly matching the effectiveness of RCC.  

\begin{figure}[!t]
\begin{center}
\includegraphics[width=0.475\textwidth,trim={0 1.2in 0 1in},clip]{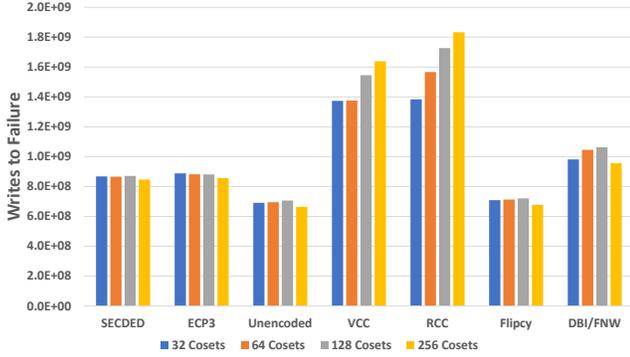}
\vspace{-0.1in}
\caption{Comparison of mean improvement in memory lifetime across benchmarks under test.}\label{fig:lifetime_mean}
\end{center}
\vspace{-0.2in}
\end{figure}
As a sensitivity study, Fig.~\ref{fig:lifetime_mean} shows the mean results of our lifetime evaluation for different coset counts.  As expected, the unencoded memory fails soonest, followed closely by Flicpy, demonstrating that this method is generally ineffective for unbiased data.  SECDED and ECP provide $\approx 20$\% improvement over unencoded.  This result is interesting in that ECP is more flexible in how it mitigates faults compared to SECDED; ECP should be able to handle several faults clustered in the same word, while SECDED would fail.  However, in practice the spatial correlation of process variation among cells means that often cells in the same row will fail early.  
DBI and FNW perform better than SECDED and ECP at $\approx 34$\% improvement over unencoded, owing to the 16-bit granularity at which they are applied.  However, considering this as a special case of 2 cosets, it fails to protect as well as the other coset approaches. VCC and RCC show the strongest results providing approximately 50-60\% and 50-64\% improvement over unencoded, respectively.  
Because VCC uses the MSB of the symbol to generate the cosets, it is slightly less flexible and this accounts for the slight benefit of RCC.  In our experience, VCC with stored cosets effectively matches RCC. 



\subsection{Performance Evaluation}
\label{sec:performance_evaluation}

\begin{figure}[tp]
\begin{center}
\includegraphics[width=0.475\textwidth]{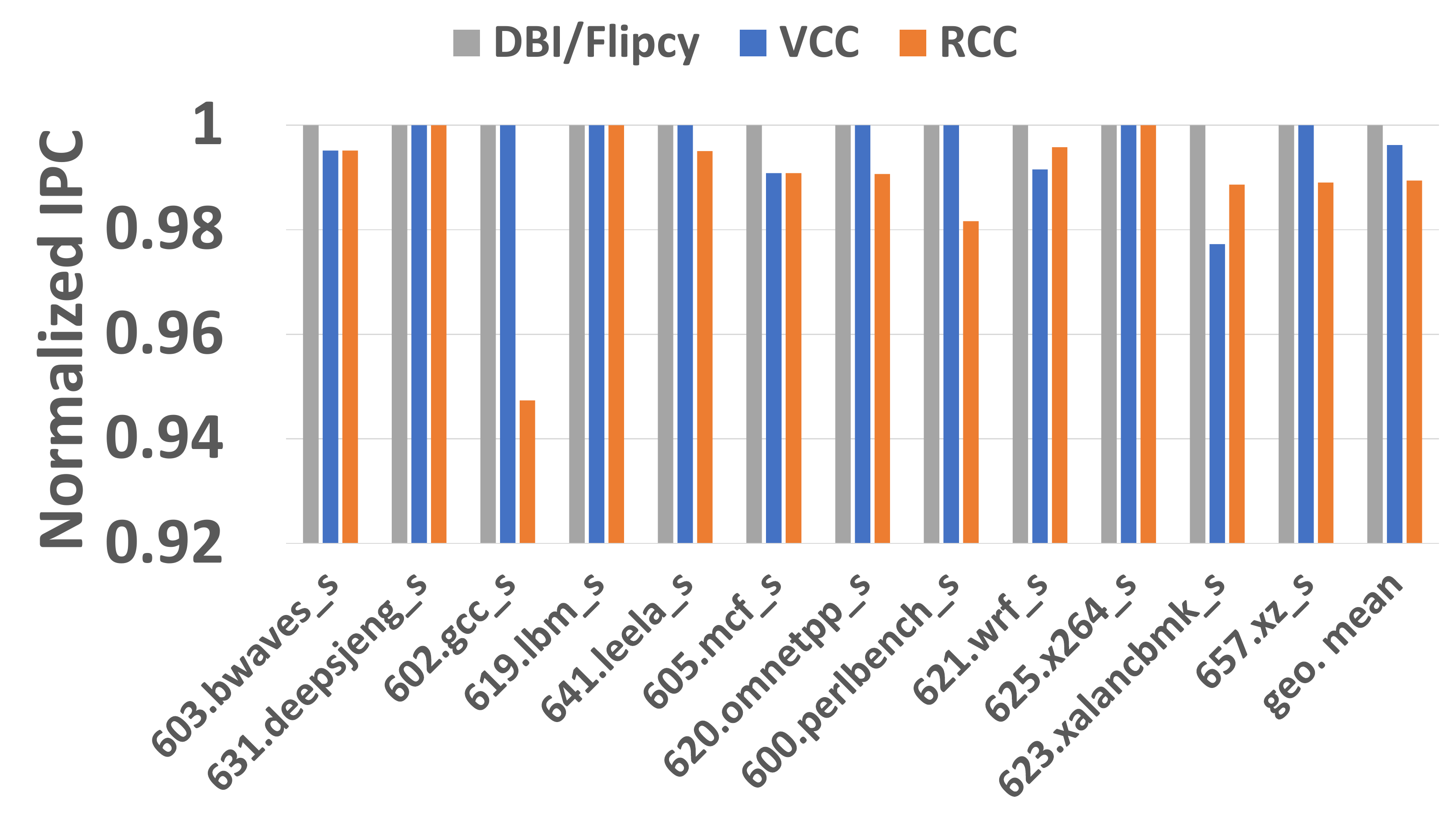}
\vspace{-0.1in}
\caption{IPC comparison across techniques normalized to unencoded writeback, using 256 coset candidates.}\label{fig:IPC}
\end{center}
\vspace{-0.2in}
\end{figure}

Fig.~\ref{fig:IPC} shows the results of our performance evaluation. 
All reported values are normalized to the IPC of the unencoded writeback case.  We can see that for DBI and Flipcy the additional delay of a few hundred $ps$ had a negligible impact on performance.  Flipcy is implemented similar to RCC except with only its three cosets, and DBI evaluates the equivalent of 2 cosets at the partition granularity.  

The performance impacts of the delay-optimized design RCC design are modest, on average $<3\%$.  This is in keeping with our hardware analysis; relative to the baseline access delay of 84ns, the additional encoding delay of 2.6ns is small.  Likewise, the 1.8ns encoding delay introduced by the VCC ($<2\%$) has a limited impact on performance on average.  In the case of VCC, we suggest that the small performance impact is justified by the energy and reliability benefits.


\section{Conclusion and Future Work}
\label{sec:conclusion}
We present Virtual Coset Coding, a technique for generating and evaluating pseudo-random codewords which dramatically increases lifetime of encrypted NVM while opportunistically reducing dynamic write energy.  Our novel technique achieves similar or better results than with random coset transformation, with the added benefit of reduced computational complexity, smaller die footprint, and more flexibility in codeword size.  Applying our technique, a memory controller can achieve 50-70\% improvement in lifetime while reducing dynamic energy up to 27\%, addressing two of the most challenging aspects of, particularly MLC, nonvolatile memories.  Compared to standard and state-of-the-art fault mitigation techniques, we achieve 36-48\% improvement in lifetime. Our technique can be further tuned to meet memory controller design constraints by adjusting the count of coset kernels and the method by which they are generated, favoring reduced area, energy budget, or encoding latency without compromising the benefits of improved lifetime and write energy reduction.  Thus, VCC takes a significant step forward towards making PCM and other NVM technologies viable for highly dense, encrypted main memory.  VCC can also be effectively applied to a system with both encrypted and non-encrypted (biased) data by adding the identity and inversion kernels. 
This naturally realizes the combined biased and random cosets similar to prior hybrid coset techniques~\cite{Hybrid_Coset,PRES}. Like prior encoding work, VCC is subject to the read-modify-write overhead, which can be mitigated through prefetching and caching~\cite{8792091}. We will explore these facets of coset encoding in future work.

\section{\textbf{Acknowledgements}}
This work was partially supported by NSF grants CNS-2133267, CNS-1822085, the Laboratory of Physical Sciences (LPS), and the NSA.
AMD, the AMD Arrow logo, and combinations thereof are trademarks of Advanced Micro Devices,
Inc. Other product names used in this publication are for identification purposes only and may be
trademarks of their respective companies.

\bibliographystyle{IEEEtranS}
\bibliography{refs}
\balance

\end{document}